\documentclass[11pt,a4paper]{article}

\usepackage{physics}
\usepackage[utf8]{inputenc}
\usepackage[T1]{fontenc}    
\usepackage{slashed}
\usepackage{mathtools}        
\usepackage{bbm}              
\usepackage{bm}               
\usepackage{caption}
\usepackage{subcaption}
\usepackage{multirow}
\usepackage{array}
\usepackage{booktabs} 
\usepackage[most]{tcolorbox}

\usetikzlibrary{positioning,arrows.meta,decorations.markings,decorations.pathmorphing}

\usepackage{jheppub}          
\usepackage{graphicx}
\usepackage{comment}
\usepackage{latexsym}
\usepackage{xspace}
\usepackage{lscape}
\usepackage{color}
\usepackage{relsize}
\usepackage{tabularx}
\usepackage{amsmath}
\usepackage{amssymb}
\usepackage{lipsum}
\usepackage{feynmp-auto}
\usepackage{upgreek}
\usepackage{nicefrac}
\usepackage[titletoc]{appendix}

\usepackage{cleveref}         
\crefname{equation}{}{}
\crefname{figure}{}{}         
\crefname{table}{}{}
\crefname{section}{}{}        
\crefname{appendix}{}{}
\crefname{footnote}{}{}

\crefalias{subequation}{equation}

\def\nc{\newcommand}

\newcommand{\be}{\begin{equation}}
\newcommand{\ee}{\end{equation}}
\newcommand{\bea}{\begin{eqnarray}}
\newcommand{\eea}{\end{eqnarray}}
\newcommand{\ba}{\begin{array}}
\newcommand{\ea}{\end{array}}
\renewcommand{\vec}[1]{\bm{#1}}

\nc{\nn}{\nonumber}

\nc{\deldag}{{\mathbin{\partial\mkern-10mu/}}}
\nc{\kdag}{{\mathbin{k\mkern-10mu /}}}
\nc{\udag}{{\mathbin{u\mkern-10mu /}}}
\nc{\Ddag}{{\mathbin{D\mkern-10mu /}}}

\def\Slashnew#1{#1\kern-0.55em\raise.05ex\hbox{/}}
\def\slashnew#1{#1\kern-0.5em\raise.05ex\hbox{{$\scriptstyle /$}}}

\def\lsim{\mathrel{\raise.3ex\hbox{$<$\kern-.75em\lower1ex\hbox{$\sim$}}}}
\def\gsim{\mathrel{\raise.3ex\hbox{$>$\kern-.75em\lower1ex\hbox{$\sim$}}}}

\nc{\shalf}{\ensuremath{\textstyle \frac{1}{2}}}
\nc{\ihalf}{\ensuremath{\textstyle \frac{i}{2}}}

\def\sfrac#1#2{\ensuremath{\textstyle \frac{#1}{#2}}}

\def\emph#1{{\em #1}}

\def\eg{{\em e.g.}}
\def\ie{{\em i.e.}}

\def\H{{\rm H}}

\nc{\sss}{\scriptscriptstyle}
\nc{\W}{{\sss W}}
\nc{\sparallel}{{\sss\parallel}}
\nc{\tGamma}{{\tilde \Gamma}}

\newcommand{\evec}[1]{{\bm{#1}}}              
\newcommand{\idx}[1]{{\rm \scriptscriptstyle{(#1)}}}

\def\H{{\rm H}}

\def\l{{\scriptscriptstyle <}}
\def\g{{\scriptscriptstyle >}}

\nc{\CPslash}{{{\scriptscriptstyle {\rm CP}}\mkern-18mu / \mkern 10mu}}

\def\calC{{\scriptscriptstyle {\cal C}}}
\def\W{{{\cal W}}}
\nc{\HF}{{\sss \rm FH}}

\title{Coherent collision integrals for neutrino transport equations}

\author[a,b]{Kimmo Kainulainen}
\author[a,b]{Harri Parkkinen}
\affiliation[a]{Department of Physics, PL 35 (YFL), 40014 University of Jyv\"askyl\"a, Finland}
\affiliation[b]{Helsinki Institute of Physics, PL 64, 00014 University of Helsinki, Finland}
\emailAdd{kimmo.kainulainen@jyu.fi}
\emailAdd{harri.h.parkkinen@jyu.fi}

\abstract{We present quantum kinetic equations for neutrinos and derive Feynman rules for computing scattering rates involving coherent states. Our rules encompass both flavour- and particle-antiparticle coherence and allow writing down the scattering matrix elements and collision integrals with the same intuitive ease as with the usual non-coherent Feynman rules. Our results are useful for computing collision rates that arise routinely in the context of coherently mixing neutrinos with arbitrary masses. We give several explicit examples, including some collision integrals for coherently mixing neutrinos in supernovae.}

\keywords{Neutrino Mixing, Neutrino Interactions, Sterile or Heavy Neutrinos}

\arxivnumber{}
\preprint{}

\begin{document}
\maketitle

%
\section{Introduction}
%

The defining property of neutrinos is their mixing and coherence. As a result, modelling neutrino evolution for example in compact objects and in early universe calls for advanced quantum kinetic equations (QKE's). Two distinct types of coherence can be identified: the flavour mixing and the particle-antiparticle mixing. QKE's describing flavour mixing have been studied extensively, but the role of the particle-antiparticle mixing is much less explored. In this paper we present QKE's that encompass both types of mixing and coherence in a unified approach. Moreover, we will derive simple generalized Feynman rules for computing the collision integrals in these QKE's, that involve interactions between coherently mixing states.

The early formulations of kinetic theory for flavour mixing neutrinos were based on the S-matrix formalism or the operator formalism~\cite{Barbieri:1989ti,Kainulainen:1990ds,Barbieri:1990vx,Enqvist:1990ad,Enqvist:1990ek,Enqvist:1991qj,Sigl:1992fn,McKellar:1992ja}. These formulations included forward scattering terms in the mean-field limit encompassing nonlinear effects induced by the neutrino-neutrino interactions. While including the flavour coherence, which can cause strong coupling between the particle and antiparticle sectors, they miss the direct couplings between the particle and antiparticle sectors through the particle-antiparticle coherence. The significance of this coherence effect to neutrino flavour evolution in hot and dense astrophysical environments has been under some discussion over the last years~\cite{Volpe:2023met}. Usually particle-antiparticle mixing is included using mean-field approximation or the BBGKY hierarchy, either neglecting collisions or including them at the ultra-relativistic (UR) limit, ignoring particle-antiparticle coherence~\cite{Volpe:2013uxl,Vaananen:2013qja, Serreau:2014cfa,Kartavtsev:2015eva}. More general transport equations were derived in~\cite{Her081,Her083,Her09,Her10,Her11,Fid11,Jukkala:2019slc, Juk21} based on the closed time path (CTP) formulation of the thermal field theory, but still without \eg~explicit treatment of dispersive corrections. In~\cite{Kainulainen:2023ocv} we  presented a fully general and self-consistent derivation of the neutrino QKE's which include particle-antiparticle mixing at the same level as the flavour mixing.

Our derivation of QKE's~\cite{Kainulainen:2023ocv} starts from Schwinger-Dyson equations in the CTP formulation~\cite{Sch61,Kel64,Cal88}, and reduces them to local QKE's assuming only the adiabaticity of background fields, validity of the weak coupling expansion and the spectral limit when computing the collision integrals and the forward scattering terms. We summarize this derivation briefly here, followed by a rigorous proof of the Feynman rules given in~\cite{Kainulainen:2023ocv} for the evaluation of collision integrals. Our formalism is valid for arbitrary neutrino masses and kinematics and encompasses both flavour and particle-antiparticle coherences. We then provide a large number of worked out examples of collision integrals to demonstrate the use of the formalism. In particular, we will derive very simple explicit expressions for the collision integrals for supernova neutrinos arising from $n\nu \leftrightarrow pe$ and $n\nu \leftrightarrow n\nu$ processes. 

This paper is organized as follows. In section~\cref{sec:QKE} we review the QKE's derived in~\cite{Kainulainen:2023ocv}. We study the emergence of energy-conserving delta functions in vertices, prove the vanishing of the overall time-dependent phase factors and present the generalized Feynman rules in section~\cref{sec:collision_integrals}. In section~\cref{sec:flavour_and_particle-antiparticle_coherence} we discuss the role of the particle-antiparticle coherence effects, showing that they usually average out in the flavour mixing scale. In sections~\cref{sec:examples} and~\cref{sec:supernova} we demonstrate the use of our formalism in different setups including supernova neutrinos. Finally in section~\cref{sec:Conclusions} we give our conclusions.

%
\section{Quantum Kinetic Equations}
\label{sec:QKE}
%

The goal of this paper is to present rules for computing scattering rates involving coherent states in the neutrino quantum kinetic equations. For this purpose we have to define precisely the QKE's including forward scattering terms and collision terms. Our presentation is very brief, but more details can be found in~\cite{Fid11,Jukkala:2019slc,Juk21} and in~\cite{Kainulainen:2023ocv}. The meaning of the coherences becomes obvious during the derivation. The basic object that holds the quantum information we seek to study, is the fermionic 2-point correlation function:
\begin{equation}
iS_{ij}(u,v) 
\equiv \Tr{\hat{\rho}_\psi \mathcal{T}_{\calC} [\psi_{i} (u) \bar{\psi}_{j} (v) ]},
\label{eq:2-point func}
\end{equation}
where $\psi$ is an arbitrary fermion field, $\hat{\rho}_\psi$ is some unknown density operator, $\mathcal{T}_{\calC}$ is time ordering operator, and $u_0,v_0$ are complex time arguments along the usual complex Keldysh-contour $\mathcal{C}$~\cite{,Kel64}. The path-ordered 2-point function $S(u,v)$ obeys the complex time contour Schwinger-Dyson equations~\cite{Cho85,Lut60, Cor74, Cal88}:
\begin{equation}
    (S_0^{-1} * S)_{\calC} (u,v) = \delta_{\calC}^\idx{4}(u - v) + (\Sigma * S)_{\calC}(u,v),
\label{eq:Schwinger}
\end{equation}
where $S_0^{-1}$ is the free inverse fermion propagator, the convolution reads $(A*B)_{\calC}(u,v) \equiv \int_{\calC} {\rm d}^4w A(u,w)B(w,v)$, and the contour time delta function is defined as $\delta_{\calC}^\idx{4}(u-v) \equiv \delta_{\calC}(u_0-v_0) \delta^\idx{3}(\vec{u}- \vec{v})$. The self-energy function $\Sigma$ depends on the model in question and can be computed \eg~from 2PI-effective action:
\begin{equation}
\label{eq:sigma var}
\Sigma_{\calC}(u,v) \equiv  -i \frac{\delta \Gamma_2[\Delta,S]}{\delta S(v,u)},
\end{equation}
where $\Gamma_2[\Delta,S]$ is the sum of the 2PI vacuum graphs of the theory, truncated to a desired order in coupling constants. We suppress the flavour and Dirac indices for simplicity whenever there is no risk of confusion.

\paragraph{Kadanoff-Baym equations}

The complex time 2-point functions and their SD-equations can be expressed in real time variables in the standard manner~\cite{Cal88}. For more details of our notations and definitions see~\cite{Her081,Her083,Jukkala:2019slc,Juk21}. In real time variables and after moving to Wigner space, one finds the Kadanoff-Baym (KB) equations, which are fully equivalent to~\cref{eq:Schwinger}:
\begin{subequations}
\label{eq:KB_fermions}
\begin{align}
\hat{\slashed{K}}S^p(k,x) - \big(\Sigma^p \otimes S^p \big)(k,x) &= 1 
\label{eq:KB_fermions_pole}
\\
\hat{\slashed{K}}S^s(k,x) - \big(\Sigma^s \otimes S^a \big)(k,x) &=
\big(\Sigma^s \otimes S^a \big)(k,x),
\label{eq:KB_fermions_stat}
\end{align}
\end{subequations}
where $\hat K = k + \sfrac{i}{2}\partial_x$, indices $p=r,a$ refer to the retarded and advanced functions and indices $s= <,>$ to the statistical Wightman functions. The Wigner space convolutions can be written as:
\begin{equation}
(\Sigma \otimes S)(k,x) \equiv e^{-\frac{i}{2}{\partial}_x^\Sigma\cdot \, \partial_k } 
               [ \Sigma_\mathrm{out}(\hat K, \, x) S(k,x) ],
\label{eq:wigconv}
\end{equation}
where $\Sigma_{\rm out}(k,x) \equiv e^{\frac{i}{2}\partial_x^{\Sigma} \cdot \, \partial_k^{\Sigma}} \Sigma(k,x)$, where the superscript $\Sigma$ indicates that the gradient $\partial^\Sigma_x$ acts only on the self-energy function, in contrast with the total derivative $\partial_k$. Note that we included the mass term into the singular part (denoted by ``sg'') of the Hermitian self-energy function: $\Sigma_\H (k,x) = \Sigma_{\textrm{H,sg}}(x) + \Sigma_{\textrm{H,nsg}}(k,x)$.

\paragraph{Local and decoupled QKE's}

Equations~\cref{eq:KB_fermions} are exact to a given approximation for the self-energy function, and their essential feature is non-locality, which is embedded in the infinite order expansions in gradients in the Wigner transformed convolution terms~\cref{eq:wigconv}. In addition the pole equations~\cref{eq:KB_fermions_pole} and the statistical equations~\cref{eq:KB_fermions_stat} are coupled to each other. To get practically useful quantum kinetic equations one must localize and decouple equations~\cref{eq:KB_fermions}. These problems were discussed carefully in~\cite{Juk21,Kainulainen:2023ocv}, where we refer the reader for details; here we give only a rough outline of the derivation. In the decoupling problem the main idea is to split the statistical function into a background part, which is strongly coupled to the pole functions, and a perturbation, whose equation can be decoupled from the set of pole- and background statistical functions with some generic assumptions for the latter. In turn, the localization problem tracks down to a truncation of the infinite order gradient expansion in the Wigner space, which can be justified by the adiabaticity assumption, or enforced by integrating over the momentum variables. The resulting local and decoupled QKE reads
\begin{equation}
    \partial_t {\bar S}_\evec{k}^\l 
    + \frac{1}{2}\{\evec{\alpha}\cdot\nabla, \, {\bar S}_\evec{k}^\l\}
    = -i\big[{\cal H}_\evec{k} ,  \bar S^\l_\evec{k} \big]
     + i \Xi^\l_\evec{k} + \bar {\cal C}^\l_{{\rm H},\evec{k}},
    \label{eq:master}
\end{equation}
where the Hamiltonian is $\smash{{\cal H}_\evec{k} = \bm{\alpha} \cdot \evec{k}\delta_{ij} + m_i\delta_{ij}\gamma^0}$, with $\smash{\alpha = \gamma^0 \gamma^i}$, and the forward scattering term $\smash{\Xi^\l_\evec{k}}$ and the Hermitian part of the collision term $\smash{\bar{\cal C}^\l_{{\rm H},\evec{k}}}$ are given in~\cite{Kainulainen:2023ocv}. The evolution of coherently mixing neutrinos is described accurately by~\cref{eq:master}, but this equation is not yet useful for practical purposes since it contains non-trivial Dirac structures.

\paragraph{Projected QKE's}

We can parametrize the Wightman functions without a loss of generality using the {\it projective representation}~\cite{Juk21,Jukkala:2019slc,Kainulainen:2023ocv}, built up using the helicity and vacuum Hamiltonian eigenbases, 
\begin{equation}
\bar S_{\evec{k} ij}^\l(t,\evec{x}) = \sum_{h a a'}  f_{\evec{k} h ij}^{\l aa'}(t,\evec{x}) 
P_{\evec{k} h i j}^{a a'},
\label{eq:correlator_par}
\end{equation}
where $\smash{f_{\evec{k} h ij}^{\l a a'}(t,\evec{x})}$ are unknown distribution functions and we defined a projection operator,
\begin{equation}
P_{\evec{k} h i j}^{a b} = N_{\evec{k}ij}^{a b} P_{\evec{k} h} P_{\evec{k} i}^a \gamma^0 P_{\evec{k} j}^{b},
\end{equation}
with the helicity and the vacuum energy projection operators given by
\begin{equation}
P_{\evec{k} h} \equiv \frac{1}{2} 
   \big( \mathbbm{1} + h \bm{\alpha} \cdot \hat{\evec{k}} \gamma^5 \big )
\qquad \textrm{and} \qquad 
P_{\evec{k} i}^{a} \equiv \frac{1}{2} 
   \big( \mathbbm{1} + a \frac{\mathcal{H}_{\evec{k} i}}{\omega_{\evec{k} i}} \big ).
\end{equation}
Here $\smash{h= \pm1}$ indicates the helicity, $\smash{a,b=\pm1}$ are energy sign indices, $i,j$ are the flavour indices and the vacuum energy of the neutrino eigenstate is defined as usual: $\omega_{\evec{k} i}= (\evec{k}^2+m_i^2)^{1/2}$. The helicity and energy projection operators satisfy the orthogonality, completeness, and idempotence relations. The normalization factors are defined as $N_{\evec{k} ij}^{a b} \equiv \sqrt{2}(1 + ab (\gamma_{\bm{k}i}^{-1} \gamma_{\bm{k}j}^{-1} - v_{{\bm k}i}v_{{\bm k}j})^{-1/2})$ with $\gamma_{\bm {k}i}^{-1} = m_i/\omega_{\bm {k}i}$ and $v_{\bm{k}i} = |{\bm k}|/\omega_{\bm{k}i}$, which leads to the standard normalization of the mass shell distribution functions in the thermal limit. 

It is now a simple task to reduce equation~\cref{eq:master} to a set of scalar equations whose solutions are characterized by their eigenfrequencies. Using the parametrization~\cref{eq:correlator_par} and multiplying~\cref{eq:master} from right by $\smash{P_{\vec{k} h j i}^{e'e}}$ and performing the trace over the Dirac indices, one finds the projected equation~\cite{Kainulainen:2023ocv}:
\begin{equation}
        \begin{aligned}
             \partial_t  f_{\vec{k} h ij}^{\l e e'} + ({\cal V}_{\bm{k} h ij}^{e'e})_{aa'} \hat{\bm k} \cdot \bm{\nabla}  f_{\vec{k} h ij}^{\l a a'} = & -2i\Delta\omega_{\bm{k} ij}^{e e'}  f_{\vec{k} h ij}^{\l e e'} 
            + \Tr[ \bar{\mathcal{C}}^\l_{\vec{k}h ij}   P_{\vec{k} h ji}^{e' e}] 
            \\
            & - i(\W^{{\rm H}ee'}_{\evec{k}hij})^{l}_{a} f_{\vec{k} h lj}^{\l a e'}
            + i[(\W^{{\rm H}e'e}_{\evec{k}hji})^{l}_{a}]^* f_{\vec{k} h il}^{\l e a},
        \label{eq:AH4}
        \end{aligned}
\end{equation}
where the repeated indices, $a$ and $l$, are summed over and  $\hat{\bm k} \equiv {\bm k}/|{\bm k}|$. The oscillation frequency is defined in terms of the frequency sign indices $e,e'$: 
\begin{equation}
2\Delta \omega_{\vec{k} ij} ^{ee'} \equiv \omega^e_{\vec{k}i} - \omega^{e'}_{\vec{k}j},
\label{eq:shell-frequencies}
\end{equation}
where $\omega^e_{{\bm k}i} \equiv e\omega_{{\bm k}i}$. The forward scattering tensor is
\begin{equation}
(\W^{{\rm H}ee'}_{\evec{k}hij})^{l}_{a}  \equiv 
\Tr[P_{\vec{k} h ji}^{e'e} \bar{\Sigma}^{\rm H}_{\vec{k}il}(\omega^a_{\bm {k}i}) P_{\vec{k} hlj}^{a e'}],
\label{eq:W-tensor}
\end{equation}
and the velocity tensor reads
\begin{equation}
({\cal V}_{\vec{k} h ij}^{e'e})_{aa'} = \delta_{a'e'} {\cal V}_{\vec{k} h ij}^{eae'}
+ \delta_{ae} {\cal V}_{\vec{k} h ji}^{a'e'e},
\label{eq:isoD-def}
\end{equation}
with
\begin{equation} 
  {\cal V}_{\vec{k} h ij}^{a b c} \equiv \frac{1}{2} N_{\vec{k} ij}^{ac} N_{\vec{k} ij}^{bc}  
  \Big( v_{{\bm k}i} \Big[ \frac{a}{(N_{\vec{k} ij}^{bc})^2} 
  + \frac{b}{(N_{\vec{k} ij}^{ac})^2}\Big] - v_{{\bm k}j}c\delta_{a,-b} \Big).
  \label{eq:Dtensor}
\end{equation}
The collision term can be written in many different ways, but an especially useful form is
\begin{equation}
  \bar{\mathcal{C}}^{\l e e'}_{\vec{k}hij} \equiv   
  \Tr[ \bar{\mathcal{C}}^\l_{\vec{k}hij} P_{\vec{k} h ji}^{e' e}] 
    = \frac{1}{2} \Big( (\W^{{\g}ee'}_{\evec{k}hij})^l_a   f_{\vec{k} h lj}^{\l a e '}
                      + [(\W^{\g e'e}_{\evec{k}hji})^l_a]^*f_{\vec{k} h il}^{\l e a } 
                      - ( > \leftrightarrow < ) \Big),
\label{eq:collision_term}
\end{equation}
where the indices $l$ and $a$ are summed over, and the ${\cal W}^{<,>}$-tensors are defined similarly to~\cref{eq:W-tensor}, with the modification $\bar \Sigma^{\rm H} \rightarrow \bar \Sigma^{\rm s}$, where $s = >,<$.

The master equation~\cref{eq:AH4} describes the neutrino evolution including flavour and particle-antiparticle coherences for arbitrary neutrino masses and kinematics and background interactions that are only assumed to be adiabatic in space. In the standard model (SM) limit and in the UR-limit it reduces to the familiar neutrino density matrix formalism; for the details see~\cite{Kainulainen:2023ocv}. All terms in~\cref{eq:AH4} have a clear physical meaning: the left-hand side is a generalized Liouville term containing a novel velocity term that describes how different group velocities affect the coherence evolution. The first term in the right hand-side is the usual Hamiltonian commutator term which encodes the relevant oscillation time scales and the second term is the collision integral that encompasses all flavour and particle-antiparticle coherences. Finally, the terms in the second row are generalized forward scattering terms.

QKE's~\cref{eq:AH4} are written in terms of frequency states rather than particle-antiparticle solutions, \ie~the Feynman-Stueckelberg interpretation has not been adapted, since this leads to a significantly simpler notation. However, by identifying positive frequency solutions as particles and negative frequency solutions with inverted 3-momenta as antiparticles, \ie~$\bar f^{\l,\g}_{\evec{k}hij} = - f^{\g,\l --}_{(-\evec{k})hij}$, it is simple to convert results between frequency solutions and the particle-antiparticle solutions.

%
\section{Collision integrals}
\label{sec:collision_integrals}
%

In this section we present Feynman rules for computing the collision integrals for coherent correlation functions that appear in our QKE's. To keep the discussion clear, we omit many details presented in~\cite{Kainulainen:2023ocv}, and showcase only the results relevant for the derivation of the Feynman rules that, in contrast, is more complete than the one given in~\cite{Kainulainen:2023ocv}.

%
\subsection{The Wightman function}
\label{sec:wightman_function}
%

While deriving the master equation~\cref{eq:AH4}, one has to deal with the decoupling problem of the KB-equations. Formally, this can be done by dividing the correlation functions into a background part and a perturbation, as discussed shortly in section~\cref{sec:QKE}. KB-equations for the perturbations admit a larger class of solutions than do the background equations: the former admits both homogeneous and inhomogeneous solutions, while the latter one have only inhomogeneous solutions. For this reason, we should write equation~\cref{eq:AH4} for the perturbation part rather than for the full Wightman function. However, in the spectral limit, which we can assume here, the background and perturbation parts can be combined, and we can write the solutions for the full Wightman function. These solutions can always be written in the form~\cite{Juk21,Kainulainen:2023ocv}
\begin{equation}
    \bar  S^s_\evec{kx} (t;u_0, v_0)  = 2 \bar{\cal A}_\evec{kx}(u_0,t)
    \,\bar S^s_\evec{kx}(t,t)\, 2 \bar{\cal A}_\evec{kx}(t, v_0),
    \label{eq:asa1}
\end{equation}
where $s = <,>$ and $\mathcal{A}_{\vec{k} x}$ is the spectral function. For the free theory the direct space spectral function reads
\begin{equation}
    \label{eq:free_spectral_function_direct}
    2 \mathcal{A}_{\vec{k} x ij}(u_0, v_0) =   e^{i\mathcal{H}_{\evec{k}i}(u_0-v_0)},
\end{equation}
where the Hamiltonian is $\smash{{\cal H}_\evec{k} = \bm{\alpha} \cdot \evec{k}\delta_{ij} + m_i\delta_{ij}\gamma^0}$ as defined earlier.
That is, $2 \mathcal{A}_{\vec{k} x ij}(u_0, v_0)$ is just the time evolution operator of the free theory, and we can rewrite the Wightman function~\cref{eq:asa1} as
\begin{equation}
         \bar S_{\evec{kx}ij}^s (t; u_0, v_0)
         = \sum_{h ab} \bar{\mathcal{S}}_{\evec{k} \evec{x} h ij}^{s ab} \; {\rm exp} \big[ 2i \Delta \omega_{\vec{k} i j}^{a b} t - i(a \omega_{\vec{k}i}u_0 - b \omega_{\vec{k}j}v_0)].
\label{eq:asa2}
\end{equation}
and we defined
\begin{equation}
    \bar{\mathcal{S}}_{\evec{k} \evec{x} hij}^{s ab} \equiv \frac{1}{2\bar\omega_{\evec{k}ij}^{ab}}
      f_{\vec{k} h i j}^{sab}(t,\evec{x}) D_{\vec{k} h i j}^{ab}, 
  \label{eq:D_definition}
\end{equation}
with 
\begin{equation}
    \label{eq:D}
        D_{\vec{k} h i j}^{ab}
        \equiv 2\bar\omega^{ab}_{\evec{k}ij} P^{ab}_{\evec{k}hij}\gamma^0 
        = ab \hat N^{ab}_{\evec{k}ij} 
        P_{\evec{k}h}(\slashed k_{i}^a+m_i)(\slashed k_{j}^b+m_j),\phantom{\Big)}
\end{equation}
where $(\smash{k_{i}^a)^\mu \equiv (\omega^a_{\evec{k}i},\evec{k})}$ and $\smash{\hat N_{\evec{k}ij}^{ab} \equiv N_{\evec{k}ij}^{ab}\bar\omega^{ab}_{\evec{k}ij}/(2\omega^a_{\evec{k}i} \omega^b_{\evec{k}j})}$. In~\cite{Kainulainen:2023ocv} we showed that~\cref{eq:D} reduces to the familiar thermal propagator in the thermal limit. The non-vanishing phase factors in~\cref{eq:asa2} for $t \neq \frac{1}{2}(u_0+v_0)$ have a crucial role in the derivation given here, by ensuring the correct energy conservation in the vertices.

\subsection{Shell projection}
\label{sec:shell_projection}

Evaluating self-energy terms in collision integrals is delicate due to the complicated coherence shell structures in correlation functions. While coherence shells are clearly separated from the particle (mass) shells, the collision integrals could not be evaluated at coherence shell frequencies without violating kinematic constraints in the collision events. This problem has been know for a long time, and in~\cite{Her081,Her082,Her083,Her09,Her10,Her11,Fid11,Jukkala:2019slc,Juk21,Juk22} it was solved by a resummation of the gradients. Indeed, the $\hat{K}$ term in the argument of the self-energy function in~\cref{eq:wigconv}, when acting on a coherence shell function, shifts the argument exactly such that the self-energy gets evaluated at a flavour-diagonal mass shell. 

We can see this also if we study the forward scattering terms appearing in the QKE~\cref{eq:AH4} starting from the direct space representation, where these terms are convolutions of a self-energy function and a propagator. When we use the form~\cref{eq:asa2} for the latter we find:
\begin{align}
    \label{eq:shell_evaluation}
     (\bar{\Sigma}^{\g } &\ast \bar{S}^\l)_{\evec{k} \evec{x}} (t,t)  = \int {\rm d}u_0 \; \bar{\Sigma}_{ \vec{k}}^\g (t,u_0) \bar{S}_{\vec{k}}^\l (u_0,t)\nonumber
    \\
     & = \sum_{h a b i j}  \int\! {\rm d} p_0 {\rm d} u_0 \, \bar{\Sigma}_{{\rm out},\vec{k}}^\g (p_0) \bar{\mathcal{S}}_{\vec{k} h i j}^{\l a b}\gamma^0 {\rm exp} \big[ 2i\Delta \omega_{\vec{k} i j}^{ab} t - i(a \omega_{\evec{k}i}u_0 \!-\! b \omega_{\evec{k}j} t) \! -\!i(t-u_0)p_0 \big]\nonumber
    \\
     &= \sum_{h a b i j}  \int {\rm d} p_0 \; \bar{\Sigma}_{{\rm out},\vec{k}}^\g (p_0) \bar{\mathcal{S}}_{\vec{k} h i j}^{\l a b} \gamma^0 \; \delta(p_0 - a\omega_{\vec{k} i}) \, {\rm exp} \; \big[ -it(p_0 - a\omega_{\vec{k} i}) \big]
     \\
     &= \sum_{h a b i j} \bar{\Sigma}_{{\rm out},\vec{k}}^\g (a\omega_{\vec{k} i}) \bar{\mathcal{S}}_{\vec{k} h i j}^{\l a b}\gamma^0\nonumber.
\end{align}
This result reproduces the object inside the trace in~\cref{eq:W-tensor}, when evaluated to the lowest order in the gradients: $\Sigma^\g_{\rm out} = \Sigma^\g$ and multiplied from the right by the projection operator $\smash{P_{\vec{k} h ji}^{e'e}}$. Note how the delta function due to the integral over the exponential phase factors in~\cref{eq:shell_evaluation} ensured that the self-energy gets evaluated at a flavour-diagonal mass shell regardless of the energy and helicity index structures. This energy conserving delta function is valid also for particle-antiparticle mixing and it also ensures that the exponential phase factor proportional to external time $t$ vanishes. We next generalize this discussion for higher order self-energy functions.

%
\subsection{Energy conservation and time-dependent phase factors}
\label{sec:energy_conservatrion}
%

When computing Feynman diagrams contributing to the forward scattering terms~\cref{eq:W-tensor} or to the collision term~\cref{eq:collision_term}, one encounters space-time integrals over phase factors associated with every internal vertex of the diagram. These integrals give rise to 4-momentum conserving delta functions as usual, although quantum coherence makes this procedure somewhat nontrivial. In spatially homogeneous and adiabatic cases, which we are interested in here, the integration over the spatial coordinates proceeds as usual, but the time-coordinate requires a more careful treatment. 

In the following discussion we assume that gauge fields are non-coherent resonances, whose propagators in the 2-time representation read as
\begin{equation}
{\cal D}_{\mu\nu}({\bm q};u_0,v_0) = \int \frac{{\rm d}q_0}{2\pi} {\cal D}_{\mu\nu}(q) e^{-iq_0(u_0-v_0)}.
\label{eq:gauge_propagator}
\end{equation}
We will also work in the 2-time representation and assume free theory limit for the pole functions, so that the phase structure of a fermion propagator is given by~\cref{eq:asa2}. Our goal is to show the emergence and the structure of the energy conserving delta-functions in individual vertices and the vanishing of the global phase proportional to $t$ for an arbitrary self-energy diagram. 

\begin{figure}
\centering
\begin{subfigure}{0.48\textwidth}
\hskip1.3cm 
\includegraphics[width=0.79\textwidth]{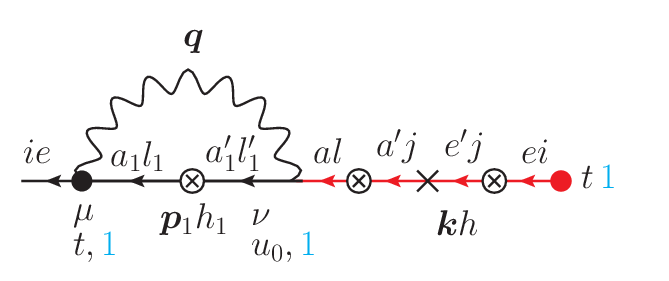}
\end{subfigure} 
\begin{subfigure}{0.38\textwidth}
    \includegraphics[width=0.84\textwidth]{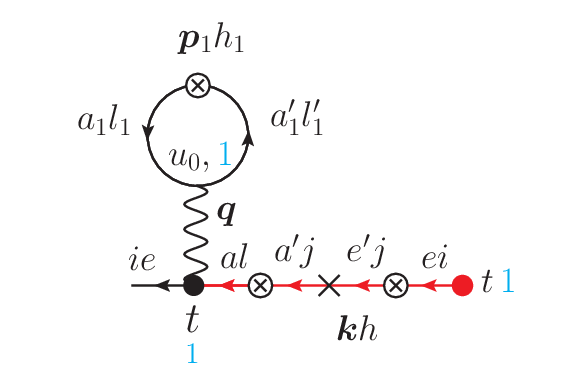}
\end{subfigure}
\caption{One-
-loop Feynman diagrams contributing to the forward scattering and collision terms. The red propagator is the dependent momentum propagator which includes the projection operator in the collision term in equation~\cref{eq:AH4}, as discussed in section~\cref{sec:Feynman_rules}.}
\label{fig:1loop_graphs}
\end{figure}

\paragraph{1-loop diagrams}

Before presenting the complete proof, we study the problem at one-loop level. This helps in introducing the notation and in identifying the basic building blocks of the proof. We
begin with the sunset diagram shown in the left panel in figure~\cref{fig:1loop_graphs}. For more details about the index structure see~\cite{Kainulainen:2023ocv}, but note that we have changed our convention from~\cite{Kainulainen:2023ocv} by reversing the direction of fermion lines in graphs, which now is in better accordance to standard practice in field theory, where one reads the fermion line against the fermion number flow\footnote{
%
%
In our old convention we graphically oriented fermion lines along the fermion number flow, but when evaluating diagrams we interpreted Feynman rules as if the line was read opposite to fermion flow. Here we simply drop this unnecessary complication by reversing the direction of fermion lines in graphs. As a result, our new vertex rules obviously follow the standard convention with respect to fermion line orientation.}. 
%
%
The black dot still shows the starting point and the red dot the ending point in the evaluation.
Using~\cref{eq:gauge_propagator} for the gauge propagator and~\cref{eq:asa2} for the fermion propagator, and introducing the usual weak interaction vertex factors, we then find
\begin{equation}
\label{eq:Z1_general_phase}
    (\mathcal{W}_{\vec{k} h ij}^{\rm{sun, H},ee'})_e^l \sim  \int {\rm d} u_0 \; e^{i\varphi_{\rm sun}(u_0) + i\alpha_{\rm sun}t} \equiv I_1[\varphi_{\rm sun}]e^{i\alpha_{\rm sun}t},
\end{equation}
where the $u_0$-independent and $t$-dependent phase factors read, respectively:
\begin{equation}
    \label{eq:Z1_phase}
    \begin{split}
        \varphi_{\rm sun}(u_0) \equiv& \,(q_0 + \omega_{\vec{p}_1 l_1'}^{a_1'}- \omega_{\vec{k} l}^{a})u_0,
        \\
       \alpha_{\rm sun} \equiv & \, 2\Delta \omega_{\vec{p}_1 l_1 l_1'}^{a_1 a_1'} + 2\Delta \omega_{\vec{k} l j}^{a e'} - \omega_{\evec{p}_1 l_1}^{a_1} +\omega_{\evec{k} j}^{e'}   - q_{0}.
    \end{split}
\end{equation}
Integrating over $u_0$ gives a delta function:
\begin{equation}
    \label{eq:Z1_delta_functions}
    I_1[\varphi_{\rm sun}] = \, 2\pi \,\delta(q_0 + \omega_{\vec{p}_1 l_1'}^{a_1'} - \omega_{\vec{k} l}^{a}),
\end{equation}
from which it is clear that the external $t$-dependent phase factor vanishes:
\begin{equation}
    \label{eq:Z1_cancellation_tphase}
    \alpha_{\rm sun} = \omega_{\vec{k} l}^{a} - \omega_{\vec{p}_1 l_1'}^{a_1'} - q_{0} = 0.
\end{equation}

The proof proceeds similarly with the tadpole diagram shown the right hand side of figure~\cref{fig:1loop_graphs}. The phase factor for this diagram can be written in the same form as~\cref{eq:Z1_general_phase}, with
\begin{equation}
    \label{eq:tadpole_phase}
    \begin{split}
        \varphi_{\rm tad}(u_0) \equiv& \, (\omega_{\vec{k} l'}^{a'}- \omega_{\vec{k} l}^{a} - q_0)u_0 ,
        \\
        \alpha_{\rm tad} \equiv & \, 2\Delta \omega_{\vec{p}_1 l_1 l_1'}^{a_1 a_1'} + q_0 = -(\omega_{\vec{k} l'}^{a'}- \omega_{\vec{k} l}^{a} - q_0).
    \end{split}
\end{equation}
Integrating over $u_0$ again produces a delta-function that makes the $t$-dependent phase coefficient $\alpha_{\rm tad}$ vanish trivially. This shows that the $t$-dependent phase factors vanish and that we obtain energy-conserving delta functions which support both flavour and particle-antiparticle mixing at least to one-loop order with gauge interactions.

\paragraph{Proof to all orders}

\begin{figure}[t]
\centering
\includegraphics[width=.8\textwidth]{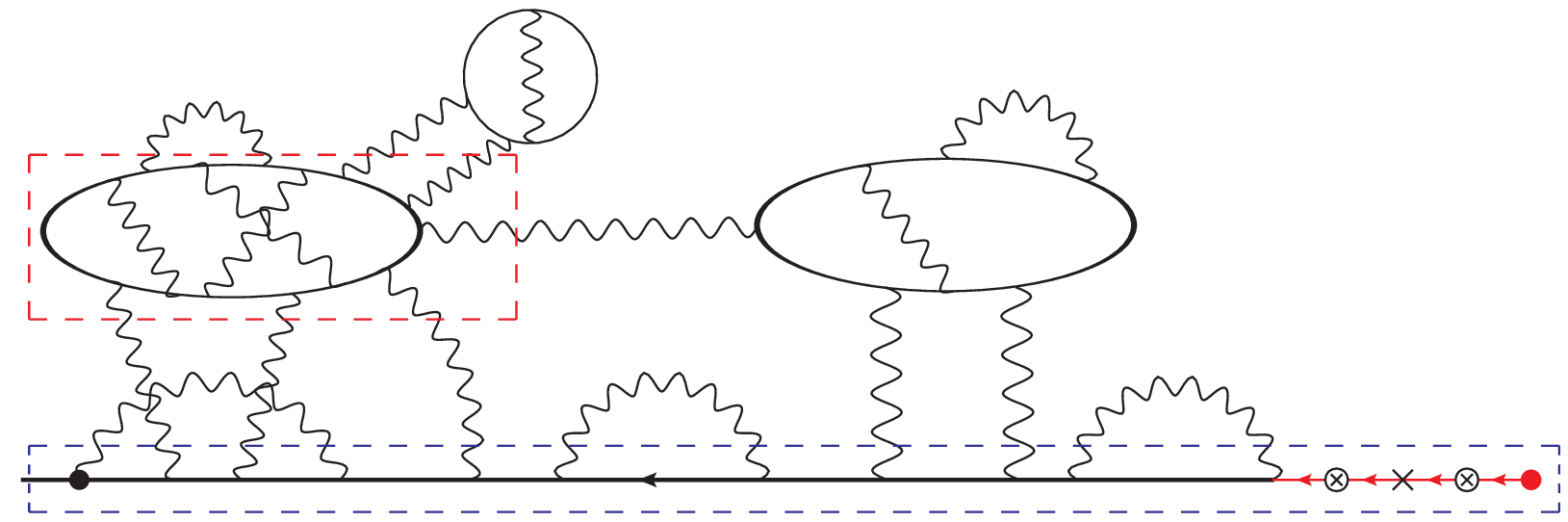}
\caption{A generic multi-loop diagram contributing to self-energy functions in a theory with gauge interactions.}
\label{fig:complicated}
\end{figure}

A general proof of vanishing of the $t$-dependent phase for an arbitrary self-energy diagram can be formulated following~\cite{Herranen:2010mh}. We show an example of a generic multi-loop diagram in figure~\cref{fig:complicated}. We only need to concentrate on two essential elements indicated by boxes in the graph: the main continuous fermion line running through the diagram, and an isolated closed fermion loop. Let us first consider the continuous line.

\begin{figure}[t]
\centering
\includegraphics[width=\textwidth]{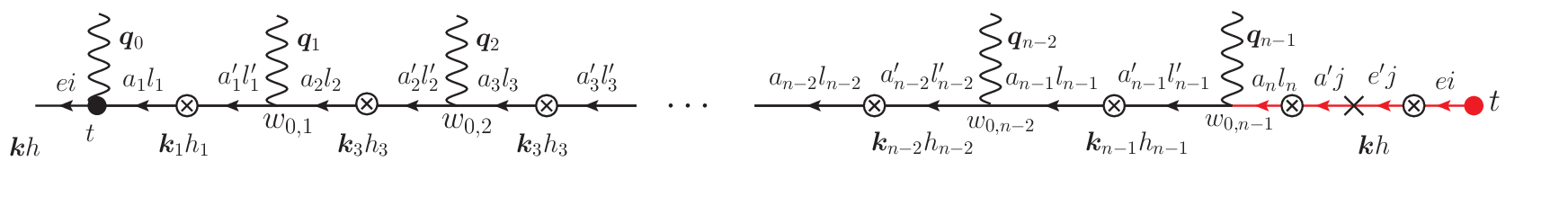}
\includegraphics[width=0.8\textwidth]{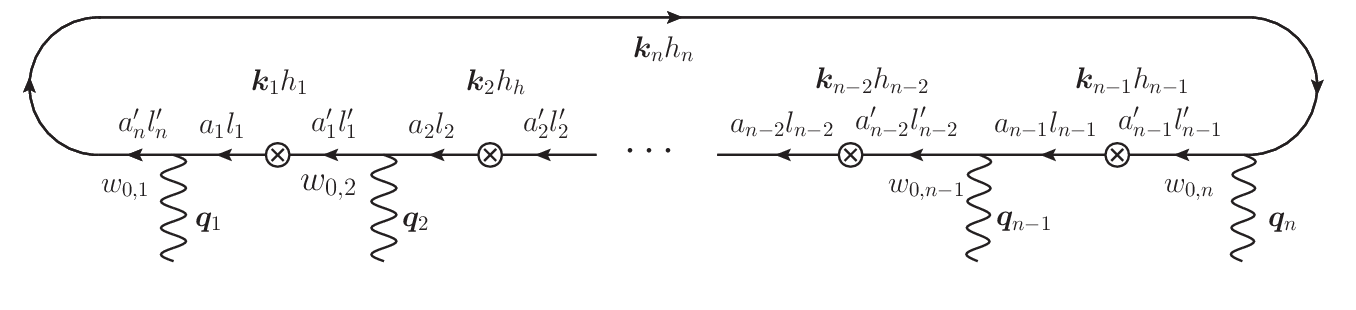}

\caption{Upper panel: the part of a generic diagram associated with the continuous fermion line shown by the dashed blue box in figure~\cref{fig:complicated}. Lower panel: a single closed loop in a generic multi-loop diagram, such as the one indicated by the red dashed box in figure~\cref{fig:complicated}.}
\label{fig:elements}
\end{figure}

We show the essential structures associated with the main fermion line in the upper panel of figure~\cref{fig:elements}. We assume that the frequency $q_{0i}$ is positive when the momentum is flowing into the diagram.  Using~\cref{eq:gauge_propagator} and~\cref{eq:asa2}, one can easily show that the total phase factor contributed by the line to the self-energy diagram is $\exp(i\Delta\varphi_{\rm line})$, where:
\begin{align}
\Delta \varphi_{\rm line} 
&= \Big( - \omega^{a_1'}_{\vec{k},l_1'} 
+ \sum_{i=2}^{n-1} 2\Delta \omega^{a_ia_i'}_{\vec{k_i}l_il_i'} 
         + \omega^{a_n}_{\vec{k}l_n} + q_{0,0}\Big) t 
+ \sum_{i=1}^{n-1} \big(q_{0,i} + \omega^{a_i'}_{\vec{k}_i,l_i'} 
                                - \omega^{a_{i+1}}_{\vec{k}_{i+1},l_{i+1}}\big) w_{0,i}
\nonumber \\
& = \Big( q_{0,0} - \sum_{i=1}^{n-1} \big( \omega^{a_i'}_{\vec{k}_i,l_i'} 
                                - \omega^{a_{i+1}}_{\vec{k}_{i+1},l_{i+1}} \big)\Big)t
+ \sum_{i=1}^{n-1} \big(q_{0,i} + \omega^{a_i'}_{\vec{k}_i,l_i'} 
                                - \omega^{a_{i+1}}_{\vec{k}_{i+1},l_{i+1}}\big) w_{0,i}
\nonumber \\
&
\rightarrow \Big( \sum_{i=0}^{n-1} q_{0,i} \Big) t,
\end{align}
where in the last line we wrote what the phase will be after the phase integrations are made over the internal times $w_{0,i}$ and the resulting delta-functions are used to convert the fermion energy differences to gauge boson frequencies. At this point we do not need to care how different vertices are connected by the propagators.

%
%

Similarly one can show that the total phase coming from a closed loop is proportional to the sum of the gauge boson frequencies flowing into the vertices in the loop. A generic loop structure is shown in the lower panel of~\cref{fig:elements}, from which we can read off the phase:
\begin{equation}
\Delta \varphi_{\rm loop} 
= \sum_{i=1}^{n} 2 \Delta \omega^{a_ia_i'}_{\vec{k_i}l_il_i'} t 
+ (q_{0,1} + \omega^{a_{n}'}_{\vec{k}_{n},l_{n}'} -\omega^{a_1'}_{\vec{k}_1,l_1'}) w_{0,1}
+ \sum_{i=2}^{n} \big(q_{0,i} + \omega^{a_{i-1}'}_{\vec{k}_{i-1},l_{i-1}'} 
                              - \omega^{a_{i}}_{\vec{k}_{i},l_{i}} \big) w_{0,i}.
\end{equation}
After integrating over the internal times $w_{0,i}$ and using the energy conservation rules, the total phase that remains can be written as
\begin{equation}
\Delta \varphi_{\rm loop} 
= -\big( \omega^{a_{n}'}_{\vec{k}_{n},l_{n}'} -\omega^{a_1}_{\vec{k}_1,l_1} \big) t
 - \sum_{i=2}^{n} \big(\omega^{a_{i-1}'}_{\vec{k}_{i-1},l_{i-1}'}- \omega^{a_i}_{\vec{k}_i,l_i} )t
= \Big( \sum_{i=1}^{n} q_{0,i} \Big)t.
\end{equation}
Summing over the phases coming from the line and all loops in the graph, one finds the total phase:
\begin{equation}
\Delta \varphi_{\rm TOT} = \Delta \varphi_{\rm line} + \sum_{\rm loops} \Delta \varphi^i_{\rm loop}
= \Big( \sum_{\rm vertices} q_{0,i} \Big)t = 0.
\end{equation}
The last sum over all vertices in the graph vanishes, because each frequency value appears in this sum exactly twice with the opposite signs, because each gauge boson propagator starts from and ends into some internal vertex in the diagram.

This completes the proof that the time-dependent phase factors cancel for all self-energy diagrams. This cancellation is essential as it allows us to write down generalized Feynman rules with {\em local} vertices and propagators for coherent particle species. 

%
%

%
\subsection{Feynman rules}
\label{sec:Feynman_rules}
%
%

After getting rid of the time-dependent phase factors in the propagators~\cref{eq:asa2} and showing how energy conserving delta functions arise, it is rather straightforward to show that the collision term can always be written as~\cite{Kainulainen:2023ocv}\footnote{We correct equation~\cref{eq:collZZ1} here slightly from~\cite{Kainulainen:2023ocv} {\em w.r.t.}~to the Hermitian conjugate term, in the computation of which one has to change $i\leftrightarrow j$ everywhere, including the $\bar\omega^{aa'}_{\vec{k}lj}$-term in front of the integral, as is evident from~\cref{eq:collision_term}.}:
\begin{equation}
    \begin{split}
    \bar{\mathcal{C}}_{\evec{k} h ij}^{\l e e'} =  \sum_{Y} \nolimits
         \frac{1}{2\bar\omega_{\evec{k}lj}^{aa'}} \int\dd{\rm{PS}_3} 
         \sfrac{1}{2}(\mathcal{M}^2)_{\evec{k}  h i j \{\evec{p}_i,Y\}}^{ee'} 
         \Lambda_{\evec{k} hj\{\evec{p}_i,Y \},x}
         +  (h.c.)^{e\leftrightarrow e'}_{i\leftrightarrow j},
    \label{eq:collZZ1}
    \end{split}
\end{equation}
where we collected all the summed indices into curly brackets, $Y \equiv \{ {\rm X}_i, h', a, a', l\}$, and defined a shorthand notation $A_{{\rm X}_i} \equiv A_{h_il_il_i'}^{a_ia_i'}$. Note how the indices are flipped in the Hermitian conjugate term. In addition, we collected all particle distribution functions into the $\Lambda$-factor: 
\begin{equation}
\label{eq:lambda factor}
\Lambda_{\evec{k} hj\{\evec{p}_i, Y \},x} = 
     f^{\l}_{{\rm X}_1\evec{p}_1}(x) \,
     f^{\g}_{{\rm X}_2\evec{p}_2}(x)\,
     f^{\l}_{{\rm X}_3\evec{p}_3}(x)\,
     f^{\g a a'}_{\evec{k} h' l j}(x) \; - > \leftrightarrow <,
\end{equation}
and defined the generalized phase space factor as
\begin{equation}
    \label{eq:phase space standard}
    \int  \dd{\rm{PS}_3} \equiv \hspace{-0.5ex}\int 
      \hspace{-0.5ex}\Big[ \prod_{i=1,3} \frac{{\rm d}^3\evec{p}_i}{(2\pi)^3 2\bar\omega_{\evec{p}_il_il_i'}} \Big] 
        (2\pi)^4 \delta^4(k^a_l + p_{2 \, l_2}^{a_2} 
        - p_{1 \, l_1'}^{a_1'}-p_{3 \, l_3'}^{a_3'}).
\end{equation}
All dynamical details of the interaction processes are included in the matrix elements squared $\smash{(\mathcal{M}^2)_{\evec{k} h ij \{\evec{p}_i Y\}}^{ee'}}$, which can be computed using the Feynman rules shown in figure~\cref{fig:feynman_fules2} and the following set of instructions:

%
\begin{figure}[t]
\centering
\includegraphics[width=0.9\textwidth]{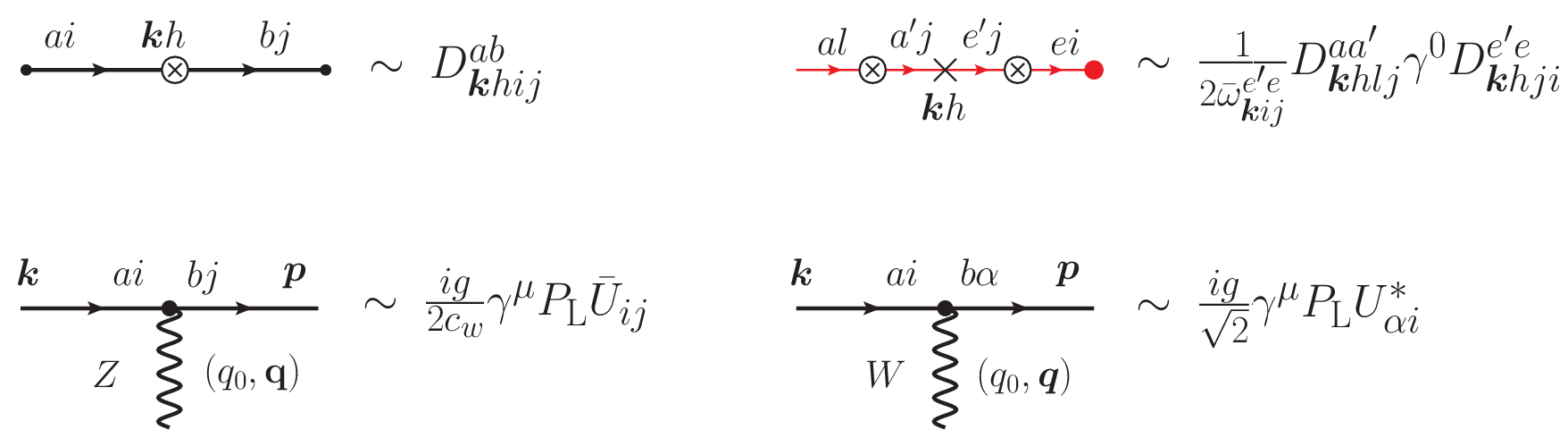}
\caption{Shown are the Feynman rules for computing matrix elements in collision integrals in the Wigner space. The first propagator should be used for all internal lines and the second, red propagator for the outgoing line. $U_{i\alpha}$ is the usual PMNS-matrix in the $W$-boson vertex, where $\alpha$ is the lepton flavour. In the $Z$-boson vertex the mixing matrix $\bar U_{ij}$ reduces to $\mathbbm{1}$ for pure active-active mixing, and $c_w = \cos\theta_w$.}
\label{fig:feynman_fules2}
\end{figure}
%

%
\begin{itemize}
\item{ Draw the loop diagrams that contribute to a given interaction process to the desired order in perturbation theory, and assign a unique momentum variable and flavour and frequency indices for each internal propagator line in the graph, allowed by the interaction vertices. For consistency, orient the direction of the main line backwards as in~\cref{fig:1loop_graphs}}.
\item{Assign the Keldysh-path indices to all vertices to isolate cuts that give rise to the desired interaction processes. You only need to evaluate $\Sigma^\g= \Sigma^{21}$ directly, so the first index is always 2 and the last 1.} 

\item{ Read off the phase space functions contributing to the $\Lambda$-factor from all internal cut propagator lines. Add the phase space factor $f_{\evec{k} hlj}^{\l aa'}/2\omega^{aa'}_{\evec{k}hlj}$, associated with the external, dependent momentum propagator (DMP), marked red in figure~\cref{fig:feynman_fules2}.}

\item{ Deduce the phase space density factor with the overall energy conserving delta function. This depends on the number of loops in the diagram and the cut one is interested in.} 
\item{Compute the matrix element squared using the Feynman rules shown in figure~\cref{fig:feynman_fules2}. Start from the equivalent of the black dot shown in the diagrams in figure~\cref{fig:1loop_graphs}, reading the line against the direction of the momentum. For each internal cut-line insert the standard propagator shown in the first diagram in~\cref{fig:feynman_fules2}. For each ("22") "11" line use the (anti) Feynman propagator. Add the DMP at the end of the fermion line it is connected to. Take a trace over the Dirac indices.}
\item{ Divide the result by two and add the Hermitian conjugate accounting for the flip of indices as indicated in~\cref{eq:collZZ1}.}
\end{itemize}
%

In~\cite{Kainulainen:2023ocv} we gave several examples of the use of these Feynman rules, including coherent neutrino-neutrino scattering including both direct and the usually neglected interference terms. This leads to the general flavour structure as discussed in section~\cref{sec:flavour_and_particle-antiparticle_coherence} and in section~\cref{sec:examples}, where we study the familiar neutrino-neutrino mixing and compare our results to literature.

%
\section{Flavour and particle antiparticle coherence}
\label{sec:flavour_and_particle-antiparticle_coherence}
%
Local coherence effects on neutrino flavour evolution have been studied actively for a long time, especially in the context of hot and dense astrophysical environments~\cite{Volpe:2023met}.
Particle-antiparticle coherence is usually studied in simplified models that assume the mean-field limit and neglect collision terms entirely, or neglect the coherence effects in them~\cite{Volpe:2013uxl,Vaananen:2013qja, Serreau:2014cfa,Kartavtsev:2015eva}. In particular it has been studied whether an instability, similar to the one known to occur due to the flavour coherence~\cite{Sawyer:2022ugt,Sawyer:2023dov,Fiorillo:2024wej}, could occur due to the particle-antiparticle coherence in forward scattering. There are also coherence effects associated with collision terms, whose treatment requires theoretical tools that did not exist until recently~\cite{Kainulainen:2023ocv}. We shall now discuss this problem using our formalism.

In Section~\cref{sec:energy_conservatrion} we showed how the overall time-dependent phase factor vanishes from the collision term. This does not mean that the collision integral is time-independent, but that its time dependence is entirely determined by distribution functions. Now, from~\cref{eq:AH4} we see that the leading time-dependent phase of a distribution function is
\begin{equation}
    \label{eq:distribution_function_phase}
    f_{\evec{k}h ii'}^{aa'} \sim -2\Delta\omega_{\evec{k}hii'}^{aa'}t.
\end{equation}
Let us now consider a generic 2-2 scattering process. Using of~\cref{eq:distribution_function_phase} in equation~\cref{eq:lambda factor} we find the following leading phase for the collision integral:
\begin{equation}
    \label{eq:time_dependence_collision_term}
    \phi_{\Lambda}(t) = - \Big(
            2\Delta\omega^{ee'}_{{\bm k}ij} 
           +2\Delta\omega^{a_1a_1'}_{{\bm p}_1l_1l_1'} 
           +2\Delta\omega^{a_2a_2'}_{{\bm p}_2l_2l_2'} 
           +2\Delta\omega^{a_3a_3'}_{{\bm p}_3l_3l_3'}\Big )t.
\end{equation}

Now consider a setup with a temporal resolution sensitive only on the flavour-oscillation time-scales ($a=a'$ in~\cref{eq:distribution_function_phase}), which are much longer than the particle-antiparticle oscillation scales ($a\neq a'$). In this case the phase $\phi_{\Lambda}(t)$ causes rapid oscillations that make the associated collision term to completely average out in the flavour scale, if any of the phase factors in~\cref{eq:time_dependence_collision_term} correspond to coherence terms ($a\neq a'$). One can make this statement precise by performing a Weierstrass transformation to coarse grain the evolution equation~\cite{Juk21,Kainulainen:2023ocv}, which indeed kills all particle-antiparticle coherences from the entire QKE~\cref{eq:AH4}%
%
%
\footnote{Weierstrass coarse graining leads to exponential suppression of the particle-antiparticle coherence contributions~\cite{Juk21,Kainulainen:2023ocv}, except in restricted part of the phase space in the collision integrals, where \eg~$\Delta\omega^{ee'}_{{\bm k}ij} + \Delta\omega^{a_1a_1'}_{{\bm p}_1l_1l_1'} \lsim \Delta m^2/\omega$. Contributions to collision integrals from these regions are suppressed by a factor $\sim {\cal O}(\Delta m^2/\omega^2)$, however.}. 
%
%
Consequently, particle-antiparticle coherences average out in most neutrino physics problems, including supernova neutrinos, even if a flavour resonance phenomenon took place there. Note that $\phi_{\Lambda}(t)$ still gives nontrivial phase evolution in the flavour scale, when the various flavour indices are not the same in~\cref{eq:time_dependence_collision_term}.

%

\begin{figure}
\centering
\includegraphics[width=0.9\textwidth]{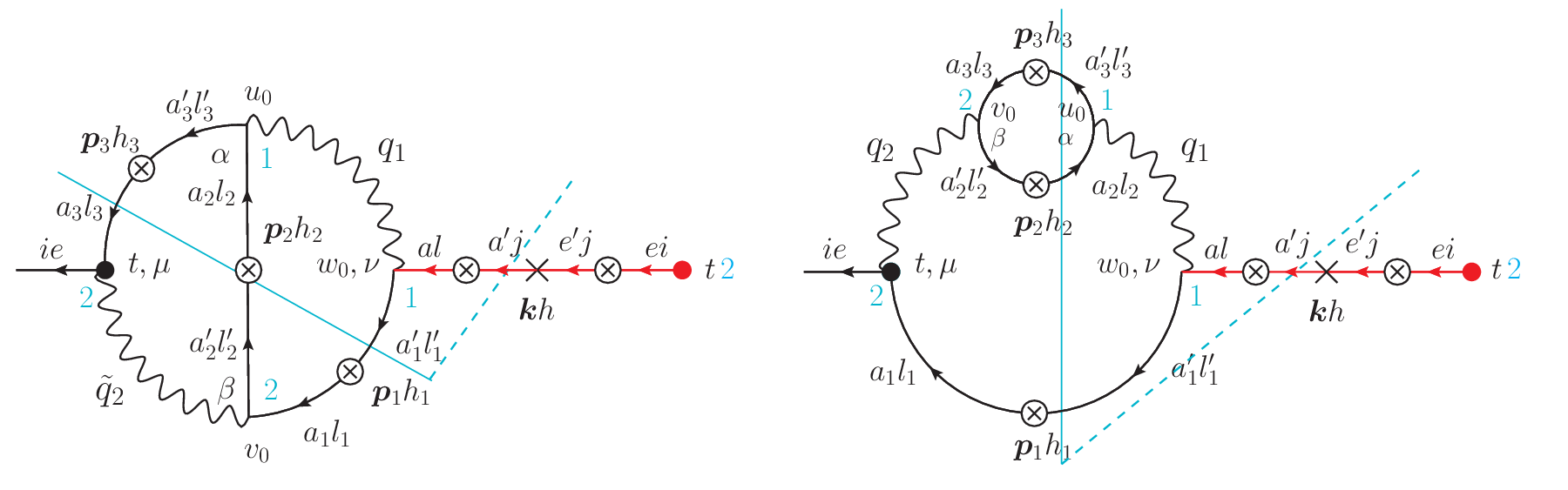}
\caption{Two-loop Feynman diagrams contributing to the 2-2 scattering terms for neutrino collision integrals mediated by neutral channel $Z$-boson interactions.}
\label{fig:2loop_graphs}
\end{figure}

Another important topic is collisions involving coherent neutrinos. In early kinetic theories neutrinos with flavour coherence are only considered to collide with flavour diagonal particles~\cite{McKellar:1992ja,Kainulainen:1990ds}. Collisions between coherent neutrino states were considered in~\cite{Sigl:1992fn,Stirner:2018ojk,Fiorillo:2024fnl}. They can be relevant in the context of the fast flavour conversion in supernovae and other hot and dense astrophysical objects, where neutrino back scattering is suggested to create an another interesting instability~\cite{Xiong:2022vsy,Lin:2022dek,PhysRevD.106.103029,PhysRevD.106.103031, Ehring:2023lcd,Tamborra:2017ubu,Tamborra:2020cul,Johns:2019izj,Capozzi:2018clo}. The most general coherent neutrino-neutrino collision terms are easily computable in our formalism. For the case of the 2-2 neutrino scatterings via neutral $Z$-boson exchange, the relevant diagrams are shown in figure~\cref{fig:2loop_graphs}. The diagram in the right gives rise to the matrix elements for the isolated $s$-, $t$- and $u$-channel processes and the diagram in left to interference terms between these channels. The associated collision integrals are easily written down using Feynman rules of section~\cref{sec:Feynman_rules}. 

%
\section{Explicit examples of collision integral evaluations}
\label{sec:examples}
%

As our first example we will reproduce the familiar results for the collision term in case of a general active-active and the two-flavour active-sterile neutrino mixing. We studied the issue briefly already in~\cite{Kainulainen:2023ocv}, but we provide more details of the derivation here and a more detailed comparison to the existing literature~\cite{Barbieri:1989ti,Kainulainen:1990ds,Barbieri:1990vx,Enqvist:1990ad,Enqvist:1990ek,Enqvist:1991qj,Sigl:1992fn, McKellar:1992ja}. The relevant Lagrangian for the system is
\begin{equation}
    \label{eq:mixing_lagrange}
    \begin{split}
        \mathcal{L}_{\rm NC} &\propto \frac{ig}{2 c_w} Z_{\mu} \bar\nu_{\alpha} \gamma^{\mu} P_{\rm L} \nu_{\alpha} + h.c.
        \\ 
        & \equiv \frac{ig}{2 c_w} \hat{U}_{\alpha \beta} Z_{\mu} \bar\nu_{\alpha} \gamma^{\mu} P_{\rm L} \nu_{\beta} + h.c.
        \\
        & = \frac{ig}{2 c_w} \bar U_{i j} Z_{\mu} \bar\nu_{i} \gamma^{\mu} P_{\rm L} \nu_{j} + h.c.,
    \end{split}
\end{equation}
where we label the vacuum basis states by Latin letters and the flavour states by Greek letters. The mixing matrix $U_{ij}$ is defined through the relation $\nu_{\alpha} = U_{\alpha i}\nu_i$, and thus $\bar U_{ij} = \sum_{\alpha\beta} U^*_{\alpha i} \hat{U}_{\alpha \beta} U_{j\beta }^* = (U^\dagger \Hat{U} U)_{ij}$. 

The 2-2 neutrino-neutrino scattering integral corresponding to the above mixing structures can be read from the diagrams in figure~\cref{fig:2loop_graphs}. In the case of light neutrinos, neglecting particle-antiparticle coherence terms, and assuming the UR-limit the collision term, becomes (for more details see~\cite{Kainulainen:2023ocv})
\begin{equation}
    \begin{split}
       \bar{\mathcal{C}}_{{\rm ZZ},\vec{k} h ij}^{\l e e}  = 
       - 16 G_{\rm F}^2  \,  \delta_{e,-h} 
         \sum_{l \{X_i\}} \frac{1}{2|\vec{k}|} & \int {\rm d PS}_3 
         (k^e \cdot p_{2}^{a_2})(p_{1}^{a_1} \cdot p_{3}^{a_3}) \; 
        \\
        & \times \left( \bar{\mathcal{U}}_{il X_i}^4 \Lambda^{aa'}_{\bm{k} hlj\{\bm{p}_i,{\rm X}_i\}}(x)  + \;  (h.c.)_{i\leftrightarrow j} \right),
       \end{split} 
    \label{eq:collZZ1_UR1xx}
\end{equation}
where the phase space factor is defined as in~\cref{eq:phase space standard}, but does not contain any flavour dependence to this order:
\begin{equation}
    \int \dd{\rm{PS}_3} = \int 
      \Big[ \prod_{i=1,3} \frac{{\rm d}^3\evec{p}_i}{(2\pi)^3 |2{\evec{p}_i}|}\; \Big] 
        (2\pi)^4 \delta^4(k^e - p_1^{a_1} + p_2^{a_2} - p_3^{a_3}).
\label{eq:phase_space_UR}
\end{equation}
All flavour dependence in the collision term is now contained in the distribution functions in the $\Lambda$-factor:
\begin{equation}
    \begin{split}
        \Lambda^{ee}_{\bm{k} lj\{\bm{p}_i,{\rm X}_i\}}(x) = f^{\g}_{X_3 {\bm p}_3}(x) f^{\l}_{X_2 {\bm p}_2}(x) f^{\g}_{ X_1 {\bm p}_1}(x) f^{\l ee}_{{\bm k}-elj}(x) -  (> \leftrightarrow <),
    \label{eq:phasefactor_again}
    \end{split}
\end{equation}
and in the mixing matrices $\bar{\mathcal{U}}_{il X_i}^4 \equiv \bar U_{il_1} \bar U_{l_1'l} \bar U_{l_2'l_3} \bar U_{l_3'l_2} + \bar U_{il_3} \bar U_{l_3'l_2} \bar U_{l_2'l_1} \bar U_{l_1'l}$. The first term in $\bar{\mathcal{U}}_{il X_i}^4$ corresponds to the direct $s$, $t$ and $u$-channel scatterings, while the second term corresponds to their interferences. 

Before moving on, we note that the collision term~\cref{eq:collZZ1_UR1xx} can also be written in terms of generalized scattering rates as follows:
\begin{equation}
    \label{eq:collZZ1_UR2}
     \bar{\mathcal{C}}_{{\rm ZZ},\vec{k} h ij}^{\l e e} = 
     - \sfrac{1}{2}\Gamma_{\evec{k} h il}^{\g e} f_{\evec{k} h lj}^{\l e} 
     + \sfrac{1}{2}\Gamma_{\evec{k} h il}^{\l e} f_{\evec{k} h lj}^{\g e} 
     - \sfrac{1}{2}f_{\evec{k} h il}^{\l e}\Gamma_{\evec{k} h lj}^{\g e *}  
     + \sfrac{1}{2}f_{\evec{k} h il}^{\g e}\Gamma_{\evec{k} h lj}^{\l e *},
\end{equation}
where
\begin{equation}
    \label{eq:gamma_tensor}
    \begin{split}
    \Gamma_{\evec{k} h il}^{\g  e} = 32 G_{\rm F}^2  \,  \delta_{e,-h}  \sum_{l \{X_i\}} \frac{1}{2|\vec{k}|} & \int {\rm d PS}_3 
        \; (k^e \cdot p_{2}^{a_2})(p_{1}^{a_1} \cdot p_{3}^{a_3}) \; 
        \\
        & \times \bar{\mathcal{U}}_{il X_i}^4 f_{\evec{p}_1 -a_1 l_1 l_1'}^{\g a_1} 
                 f_{\evec{p}_2 -a_2 l_2 l_2'}^{\l a_2} 
                 f_{\evec{p}_3 -a_3 l_3 l_3'}^{\g a_3}.
        \end{split}
\end{equation}
The rate $\Gamma_{\evec{k} h il}^{\l e}$ is obtained from equation~\cref{eq:gamma_tensor} by replacing $< \; \leftrightarrow \; >$ in the distribution functions. The collision integrals~\cref{eq:collZZ1_UR1xx} and~\cref{eq:collZZ1_UR2} are completely general expressions valid for interactions of coherent light neutrinos with arbitrary flavour mixing structure.

Collision integrals are often expressed in the flavour basis rather than the (vacuum) mass basis. To this end we need to rotate equation~\cref{eq:collZZ1_UR2} to the flavour basis:
\begin{equation}
    \label{eq:rotation_of_collision_term1}
     \bar{\mathcal{C}}_{\alpha \beta}= (U \bar C U^{\dagger})_{\alpha \beta} \sim (U \, \Gamma f U^\dagger)_{\alpha \beta} = (U_{\alpha i} \, \Gamma_{il} U_{\delta l}^{*}) (U_{\delta l} f_{lj} U_{\beta j}^{*}) = \Gamma_{\alpha\delta}f_{\delta\beta}.
\end{equation}
Rotating the collision rate $\Gamma_{\delta\beta}$ has some complicated flavour dependence in the distribution functions contracted with the mixing matrix tensor $\bar{\mathcal{U}}_{il X_i}^4$. After some algebra one finds:
\begin{align}
    \label{eq:rotation_of_distributions}
    \Gamma^\g_{\alpha\delta} &\sim U_{\alpha i} \; \bar{\mathcal{U}}_{il X_i}^4 
                f_{\evec{p}_1 -a_1 l_1 l_1'}^{\g a_1} 
                 f_{\evec{p}_2 -a_2 l_2 l_2'}^{\l a_2} 
                 f_{\evec{p}_3 -a_3 l_3 l_3'}^{\g a_3} \; U^*_{\delta l} 
     \nonumber \\
     &= (\hat{U} \hat f_{\evec{p}_1 -a_1}^{\g a_1} \hat{U})_{\alpha\delta} \;
        { \rm Tr} \Big[\hat{U} \hat f_{\evec{p}_2 -a_2}^{\l a_2} \hat{U} 
                               \hat f_{\evec{p}_3 -a_3}^{\g  a_3}\Big]
      + (\hat{U} \hat f_{\evec{p}_3 -a_3}^{\g a_3} \hat{U} \hat f_{\evec{p}_2 -a_2}^{\l a_2} \hat{U}
     \hat f_{\evec{p}_1 -a_1}^{\g a_1} \hat{U})_{\alpha \delta},
\end{align}    
where the first term in the right hand side arises from the direct terms and the second term from the interference terms.  All objects in the right hand side are matrices in the flavour basis, which we have emphasized by the use of the hat-symbol. To proceed to specific examples beyond these general results, we need to specify the rotation matrices $\hat U_{\alpha\beta}$.

%
\paragraph{Two-flavour active-sterile mixing}
%

Consider first the two-flavour active -- sterile mixing. In this case the mixing matrices read
\begin{equation}
\label{eq:mixing_matrices_sterile}
U^{\rm as}_{\alpha i} = \left( \begin{array}{cc}
                   c &s \\
                   -s & c 
                   \end{array}\right),
\qquad
\hat U^{\rm as}_{\alpha\beta} = \left( \begin{array}{cc}
                 1 & 0 \\
                 0 & 0 
                   \end{array}\right)
\quad \Rightarrow \quad
\bar U^{\rm as}_{ij} = \left( \begin{array}{cc}
                 c^2 & cs \\
                 cs & s^2 
                   \end{array}\right),
\end{equation}
where \eg~$c \equiv \cos{\theta}$, where $\theta$ is the vacuum mixing angle. Using this $\hat U^{\rm as}_{\alpha\beta}$ in equation~\cref{eq:rotation_of_distributions} and then plugging the result into~\cref{eq:rotation_of_collision_term1}, one readily obtains
\begin{equation}
\label{eq:active_sterile_coll}
    C^{ee}_{{\rm ZZ}\bm{k}h} 
      = -\left( \begin{array}{cc}
              \;\;\Gamma^{\g e}_{{\rm ZZ}\bm{k}h aa} f^{\l e}_{\bm{k}haa} &
              \sfrac{1}{2}\Gamma^{\g e}_{{\rm ZZ}\bm{k}h aa}\, f^{\l e}_{\bm{k}has} \\
              \sfrac{1}{2}\Gamma^{\g e}_{{\rm ZZ}\bm{k}h aa}\, f^{\l e}_{\bm{k}hsa}& 0 
        \end{array}\right) - \big( < \leftrightarrow >\big),
\end{equation}
where the real valued purely active collision rate reads
\begin{equation}
       \Gamma^{\g e}_{{\rm ZZ}\bm{k}h aa} \equiv
         32 G_{\rm F}^2  \, \delta_{e,-h} 
        \sum_{\{a_i\}} \frac{1}{2|\vec{k}|}  \int {\rm d PS}_3  
          \; (k^e \cdot p_{2}^{a_2})(p_{1}^{a_1} \cdot p_{3}^{a_3}) 
          \; f^{\g a_1}_{h_1\evec{p}_1aa}
             f^{\l a_2}_{h_2\evec{p}_2aa}
             f^{\g a_3}_{h_3\evec{p}_3aa}.
    \label{eq:raterate}
\end{equation}
In the calculation we used $(f^{\l e}_{\bm{k}hsa})^* = f^{\l e}_{\bm{k}has}$ in the Hermitian conjugate term. The rate~\cref{eq:raterate} also contains equal contributions from the direct term and from the interference term which are summed together. In~\cref{eq:active_sterile_coll} sterile state has no collision integral in the flavour basis, and the off-diagonal terms are suppressed by a factor of one-half when compared to the active rate. All results presented here contain both particle and antiparticle channels. While the latter are described by negative-frequency solutions they can be easily connected to the antiparticles using the Feynman-Stueckelberg rule $\smash{\bar f^{\l,\g}_{\evec{k}hij} = - f^{\g,\l -}_{(-\evec{k})hij}}$ and the general identity $f^{\g ab}_{{\bm p}hij} = a\delta_{ij}\delta_{ab}-f^{\g ab}_{{\bm p}hij}$.

%
\paragraph{Active-active mixing}
%
Consider now the arbitrary dimensional active-active flavour mixing, which corresponds to $\hat{U}_{\alpha \beta} = \delta_{\alpha \beta}$. As opposed to the active-sterile mixing, we have to now specify also the interaction channel in order to proceed further. We choose to consider the neutrino-neutrino scattering, \ie~we set $e=a_1=a_2=a_3=1$. We continue to assume the UR-limit and furthermore the Maxwell-Boltzmann limit: $f_{\evec{k}h ij}^{\g  e} = \delta_{ij}$. Using again~\cref{eq:rotation_of_collision_term1} and~\cref{eq:rotation_of_distributions}, one can write the collision integral as a flavour basis matrix
\begin{align}
    \label{eq:active_flavour_collision_term_2}
        \bar{\mathcal{C}}_{{\rm ZZ},\vec{k} h}^{\l}  = &- 16 G_{\rm F}^2  \,  \delta_{h, -1} \frac{1}{2|\vec{k}|} \int {\rm d PS}_3 
        \; (k \cdot p_{2})(p_{1} \cdot p_{3}) 
        \nonumber\\
        & \times\Big( 2 \Tr[\hat f_{\evec{p}_2h}^{\l}] \hat f_{\evec{k} h}^{\l} 
         - 2 \Tr[\hat f_{\evec{p}_3 h}^{\l}] \hat f_{\evec{p}_1 h}^{\l} 
        \\
        & + \hat f_{\evec{p}_2 h}^{\l} \hat f_{\evec{k}   h}^{\l}
          + \hat f_{\evec{k}   h}^{\l} \hat f_{\evec{p}_2 h}^{\l} 
          - \hat f_{\evec{p}_3 h}^{\l} \hat f_{\evec{p}_1 h}^{\l} 
          - \hat f_{\evec{p}_1 h}^{\l} \hat f_{\evec{p}_3 h}^{\l} \Big),
        \nonumber
\end{align}
where all $\hat f$'s on the right are (density) matrices in the flavour space. The first two terms, which include traces of distribution functions, arise from the direct graphs, and the rest of the terms are contributions from the interference terms. The latter drop out if one considers scattering between non-mixing species. If we now assume this limit, and write $\smash{{\rm Tr}[f_{\evec{p}_2 -1}^{\l}] \rightarrow \sum_j n_j(p_2)}$, the resulting collision integral of a coherent state off a number of different, non-mixing particle species, reads
\begin{align}
    \label{eq:collision_term_McK}
    \overline{\mathcal{C}}_{{\rm ZZ},\vec{k} h}^{\l} 
    = - 32 G_{\rm F}^2  \, \delta_{h, -1} \frac{1}{2|\vec{k}|} 
    \sum_j \int {\rm d PS}_3 \; (k \cdot p_{2})(p_{1} \cdot p_{3}) 
    \,\big( n_j(p_2) \hat f_{\evec{k}h    }^\l 
                - n_j(p_3) \hat f_{\evec{p}_1 h }^\l \big),
\end{align}
where $n_j(p)$ are scalar valued particle distribution functions. This result is identical to the one obtained in~\cite{McKellar:1992ja} for this case. From the above discussion it is evident what simplifying assumptions are necessary to recover the known results as limiting cases of our formalism, and how it generalizes the usual kinetic theory formalism.

%
\section{Supernova neutrinos}
\label{sec:supernova}
%

As our last example, we consider the light neutrino transport and scattering in supernovae. In this case we can assume the UR-limit and neglect the particle-antiparticle coherence, as discussed in section~\cref{sec:flavour_and_particle-antiparticle_coherence}. The master equation~\cref{eq:AH4} then reduces to a particularly simple form~\cite{Kainulainen:2023ocv}\footnote{Our discussion obviously neglects the space-time curvature. Curvature could be included in the derivation along the lines as was done for the expanding Friedman-Robertson-Walker background in~\cite{Juk21}. In the UR-limit it could be included also by generalizing the Liouville term as in~\cite{Richers:2019grc}.}:
\begin{equation}
\partial_t  f_{\vec{k} h ij}^{e} 
           + \bar{v}_{{\bm k}ij}\, \hat{\bm k} \cdot \bm{\nabla} f_{\vec{k} h ij}^{e} =
- i[H_{\evec{k}h}^e, f_{\vec{k} h}^{e}]_{ij} + \bar{\mathcal{C}}_{\vec{k} hij}^{\l  e e'},\strut
\label{eq:AH_UR}
\end{equation}
where $\bar{v}_{{\bm k}ij} \equiv \sfrac{1}{2}(v_{{\bm k}i} + v_{{\bm k}j} )$ is the average of the velocities of the flavour states $i$ and $j$ with momentum ${\bm k}$, the collision term is defined in~\cref{eq:collZZ1}, and the Hamiltonian reads
\begin{equation}
(H^e_{\evec{k}h})_{ij} = e \delta_{ij}\omega_{\evec{k}i} + (V^e_{\evec{k}h})_{ij}.
\label{eq:matter_hamiltonian}
\end{equation}
The forward scattering potential is easily computed following~\cite{Kainulainen:2023ocv} and is given by
\begin{equation}
\label{eq:V-components}
\begin{split}
(V^{e}_{\evec{k}h})_{il} = \delta_{e,-h}\sqrt{2}G_{\rm F}\Big\{ \, 
& U^*_{ei}U_{ej} \Delta n_e 
 + \frac{1}{2}\bar U_{il} \!\!\! \sum_{a=e,p,n} \!\! v^{\rm\scriptscriptstyle Z}_a \Delta n_a  
\\
&+ (\bar U \Delta n_{\nu\evec{k}} \bar U)_{il} + \bar U_{il}{\rm Tr}[\bar U \Delta n_{\nu\evec{k}}]\,\Big\},
\end{split}
\end{equation}
where $\Delta n_a = n_a-\bar n_a$, where $n_a(t,{\vec x})$ are the number densities including internal degrees of freedom, and $v^{\rm \scriptscriptstyle Z}_a$ are the $Z$-boson vector couplings. The asymmetry matrices $\Delta n_{\nu\evec{k}ij}$ in the last two terms in~\cref{eq:V-components}, which come from neutrino-neutrino scattering are given by
\begin{equation}
\Delta n_{\nu\evec{k}ij}(t,\evec{x}) \equiv
\int \frac{{\rm d}^3\bm{q}}{(2\pi )^3} (1-\hat {\bm q}\cdot \hat {\bm k})
\big( f_{\vec{q}-ij}^\l(t,\evec{x}) - \bar f_{\vec{q}+ij}^\l(t,\evec{x})\big).
\label{eq:WH_spectral_limit2}
\end{equation}
In the Standard Model the vector couplings for electrons, protons and neutrons are $\smash{v^{\rm\scriptscriptstyle Z}_e} = -1 + 4\sin\theta_W$, $\smash{v^{\rm\scriptscriptstyle Z}_p} = 1 - 4\sin\theta_W$ and $\smash{v^{\rm\scriptscriptstyle Z}_n} = -1$, respectively
Because proton and electron densities are expected to be equal due to charge neutrality, their contributions actually cancel in the sum term in~\cref{eq:V-components}. Also, in the pure active-active mixing limit, where $\bar U_{ij}\rightarrow \mathbbm{1}$, all neutral current induced tadpole terms (the last terms in both lines in~\cref{eq:V-components}) become ineffective in the commutator in~\cref{eq:AH_UR} and can be neglected.

The second term in~\cref{eq:AH_UR} holds information of how different group velocities affect the coherence evolution of neutrinos. This information was missing in the QKE's of~\cite{Richers:2019grc} based on the $\epsilon$-expansion of~\cite{Vlasenko:2013fja}, but was included in refs.~\cite{Stirner:2018ojk,Fiorillo:2024fnl}. Inside supernova the effect of velocity differences are negligible due to short propagation distances and one can safely set $\smash{v_{{\bm k}ij} \rightarrow 1}$ there. However, during the propagation from supernova to Earth, the velocity differences can cause spatial separation between different flavour density matrix components, making them arrive to Earth at different times, which could affect the detected neutrino fluxes. In particular it is possible that the coherence solutions would completely separate from the diagonal solutions reducing the latter to pure statistical limit. This raises an interesting question: how would the pure coherence pulse interact with a detector? One can answer this and other questions about the neutrino detection by computing the relevant collision terms between the neutrino density matrix describing neutrinos and the detector atoms using our Feynman rules. We leave these questions for a future work however, and concentrate on the scattering of neutrinos in the supernovae.  

\begin{figure}[t]
\qquad\quad
\centering
\begin{subfigure}{.45\textwidth}
\includegraphics[width=.8\textwidth]{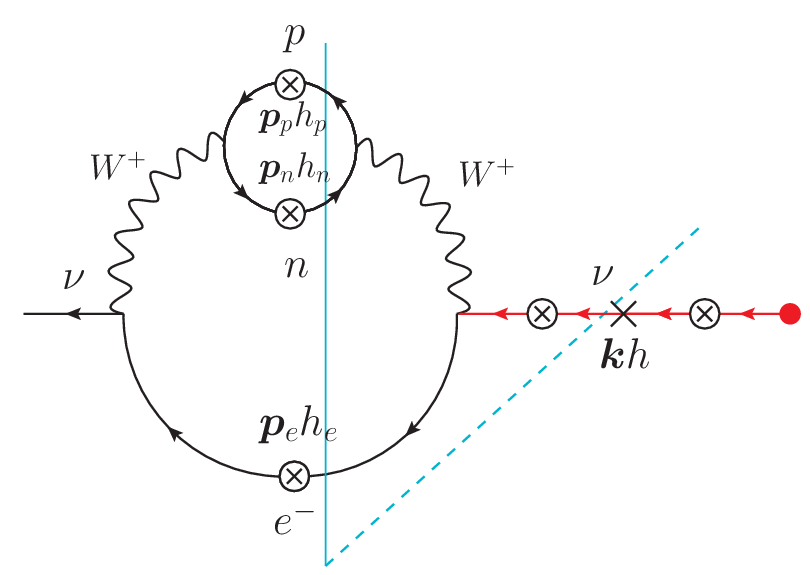}
\end{subfigure}
\begin{subfigure}{.45\textwidth}
  \includegraphics[width=.8\textwidth]{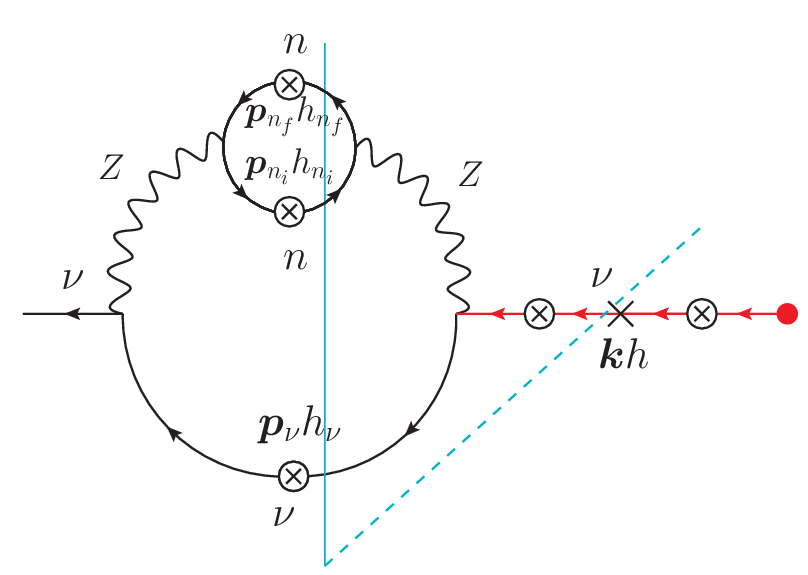}
\end{subfigure} 
\caption{Shown are graphs and cuts from which the considered neutrino scattering and neutrino production channels arise for neutron stars.}
\label{fig:supernova}
\end{figure}
%

%
\subsection{Scattering processes in supernovae}
%

The relevant neutrino scattering channels in supernovae are neutrino absorption and emission, $\smash{p + e \leftrightarrow \nu_e + n}$, neutrino-nucleon scattering $\smash{\nu + n \rightarrow \nu + n}$, neutrino scattering off electrons, electron-positron pair annihilation to neutrinos and neutrino-neutrino scatterings. In the previous section we already presented a collision integrals relevant for neutrino-neutrino scattering~\cref{eq:active_flavour_collision_term_2}. Also, the collision term for the neutrino-scattering off electrons is eventually equivalent with~\cref{eq:collision_term_McK} up to a constant. Instead of presenting a full analysis of all contributing terms, we concentrate on the neutrino absorption and emission and the neutrino-nucleon scattering processes here. For a more exhaustive analysis of neutrino scatterings in supernovae in an alternative QKE setup, see~\cite{Richers:2019grc}.
 
\paragraph{Matrix elements for the absorption and nucleon scattering processes}

The graphs and cuts corresponding to the $pe$ and $n\nu$ processes are shown in figure~\cref{fig:supernova}.  In the UR-limit we can set the $Z$ and $W$ boson propagators to $\smash{D^{\rm Z/W}_{\mu\nu} = g_{\mu \nu}/M_{\rm Z/W}^2}$. Furthermore, for the vector boson interactions with protons and neutrons we use vertex factors:
\begin{equation}
Znn \sim \frac{ig}{4\cos \theta_W} (g_V^{\scriptscriptstyle \rm Z} - g_A^{\scriptscriptstyle \rm Z}\gamma^5) 
\quad {\rm  and} \quad 
Wpn \sim  \frac{ig}{2\sqrt{2}}(g_V^{\scriptscriptstyle \rm W}-g_A^{\scriptscriptstyle \rm W}\gamma^5),
\end{equation}
where $g^{\scriptscriptstyle\rm X}_{V,A}$ are the appropriate structure functions. Following the instructions presented in section~\cref{sec:Feynman_rules}, it is then straightforward to show that the matrix elements squared for the $pe$-process, coming from the first graph in figure~\cref{fig:supernova}, is:
\def\gva#1#2{{g_{\scriptscriptstyle{#1}}^{\rm \scriptscriptstyle #2}}}
\begin{align}
        \label{eq:matrix_element_supernova_pe}
         (\mathcal{M}^2)_{\evec{k}ij}^{n\nu-ep}(p_1,p_2,p_3) &\approx 32 G_F^2 U^*_{e i} U_{e l}
         \Big( 
          (\gva{V}{W} + \gva{A}{W})^2  (k \cdot p_{n}) (p_{e} \cdot p_{p}) 
         \nonumber \\
         & \hskip 2.35cm
         + (\gva{V}{W} - \gva{A}{W})^2  (k \cdot p_{p}) (p_{e} \cdot p_{n}) 
         \nonumber \\
         & \hskip 1.92cm
         - [(\gva{V}{W})^2 - (\gva{A}{W})^2 ] m_n m_p (k \cdot p_{e}) \,
         \Big)
         \nonumber \\
         & \equiv 32 G_F^2 U^*_{e i} U_{e l} m_n m_p {\rm F}_{W}(k,p_e,p_n,p_p),
\end{align}
where $G_F$ is the Fermi constant,  $m_n$ and $m_p$ are the neutron and proton masses and $U_{i\alpha}$ is the usual PMNS-matrix. The matrix element for the $n\nu$-process has the same kinematic structure:
\begin{equation}
    \label{eq:matrix_element_supernova_nnu}
    (\mathcal{M}^2)_{\evec{k}ijl_1l_1'}^{ee}(p_1,p_2,p_3) 
    \approx -8 G_F^2 \bar U_{i l_1}\bar U_{l_1' l} m_n^2 {\rm F}_{Z}(k,p_\nu,p_{n_i},p_{n_f}),
\end{equation}
where $\bar U_{i l_1} = (U^\dagger \Hat{U} U)_{ij}$ as defined in the previous section. The matrix elements squared functions~\cref{eq:matrix_element_supernova_pe,eq:matrix_element_supernova_nnu} can be simplified in the case of a supernova environment, where $m_e \ll T \ll m_p, m_n$. One can then assume that the initial state neutrons are at rest, the nucleon momenta are negligible compared to their masses, and we can neglect both the electron and neutrino masses. In this limit one finds: 
\begin{equation}
\label{eq:sn_matrix_elements}
    \begin{split}
  &{\rm F}_{W}(k,p_e,p_n,p_p)
     \approx  |{\vec k}| |{\vec p_e}| b_{\rm W} (1 - a_{\rm W} \hat {\vec k} \cdot \hat {\vec p}_e),
 \nonumber\\
  &{\rm F}_{Z}(k,p_e,p_{n_i},p_{n_f})
     \approx  |{\vec k}| |{\vec p_\nu}| b_{\rm Z} (1 - a_{\rm Z} \hat {\vec k} \cdot \hat {\vec p}_{\nu}),
     \end{split}
\end{equation}
where
\begin{equation}
b_G = (\gva{V}{G})^2 + 3(\gva{A}{G})^2 \quad {\rm and} \quad  a_G = ((\gva{V}{G})^2 - (\gva{A}{G})^2)/b_G,
\end{equation}
with $G=W,Z$. 

%
\subsection{Neutrino absorption and emission}
%

Together with the phase space element~\cref{eq:phase space standard} and the distribution function~\cref{eq:lambda factor} the matrix elements~\cref{eq:sn_matrix_elements} allow us to write the collision integrals in rather simple forms. For the neutrino absorption and emission process we find
\begin{align}
\bar C^{\l,pe}_{{\vec k}hij} \approx -16 G_{\rm F}^2  \, \delta_{h, -1} \frac{1}{2|\vec{k}|} 
&\int {\rm dPS}_3 (2\pi )^4\delta^4(k-p_e+p_n-p_p) 
\nonumber\\
& \times  \Big(U_{ei}^* U_{e l} m_n^2 |{\vec k}| |{\vec p_e}| 
           b_{\rm W} (1 - a_{\rm W} \hat {\vec k} \cdot \hat {\vec p}_e)
\nonumber\\
&\phantom{Ha}\times (f^{\g}_{{\bm p}_p} f^{\l}_{{\bm p}_n} f^{\g}_{{\bm p}_e} f^{\l}_{{\bm k}hlj} -  
f^{\l}_{{\bm p}_p} f^{\g}_{{\bm p}_n} f^{\l}_{{\bm p}_e} f^{\g}_{{\bm k}hlj} \big) 
+ (h.c.)_{i\leftrightarrow j} \Big).
\label{eq:pe_first_version}
\end{align}
We can assume that distributions for protons, neutrons and electrons are in local thermal equilibrium. We also assume the MB-limit and note that due to isotropy of the electron distribution the $\hat{\vec k}\cdot \hat{\vec p}_e$-term vanishes, so that eventually
\begin{equation}
\bar C^{\l,pe}_{{\vec k}hij} = -\frac{b_W}{2\pi} G_{\rm F}^2 n_n (|{\vec k}|+\Delta m)^2 \delta_{h,-1}
\left( U^*_{ei}U_{el}(f^\l_{{\vec k}hlj} - \delta_{lj}f^{{\rm eq}}_{{\vec k}jj}) + (h.c.)_{i \leftrightarrow j}\right),
\label{eq:pe_second_version}
\end{equation}
where $\Delta m \equiv m_n-m_p \approx 1.29$ MeV, $n_n$ is the neutron number density and $f^{{\rm eq}}_{{\vec k}jj}$ is the diagonal Maxwell-Boltzmann equilibrium density. It is instructive to write down also the flavour basis version of this collision integral. One easily finds
\begin{equation}
\bar C^{\l,pe}_{{\vec k}h\alpha\beta} 
= - \frac{b_W}{2\pi} G_{\rm F}^2 n_n (|{\vec k}|+\Delta m)^2 \delta_{h,-1}
\big( \delta_{\alpha e} \hat f^\l_{{\vec k}he\beta  } 
    + \delta_{e \beta } \hat f^\l_{{\vec k}h\alpha e} 
    - 2\delta_{\alpha e}\delta_{e \beta} \hat f^{\rm eq}_{{\vec k}} \big),
\label{eq:pe_third_version}
\end{equation}
where $\hat f_{\alpha\beta} = (U)_{\alpha i}f_{ij} U^*_{\beta j}$. Because thermal equilibrium is defined by the Hamiltonian eigenstates, equilibrium distributions in general contain off-diagonal components in flavour basis. These terms are of order $\Delta m_{ij}^2/T^2$ however, and one can set ${\hat f}^{\rm eq}_{{\vec k}h\alpha\beta} = \delta_{\alpha\beta} f^{\rm eq}_{\vec k}$ to an excellent accuracy in supernovae, where $f^{\rm eq}_{\vec k}$ is the massless MB-distribution. This is what we did in~\cref{eq:pe_third_version}. Note also that the diagonal $ee$-component of the density matrix $\hat f^\l_{{\vec k}hee}$ is brought to equilibrium with the rate 
\begin{equation}
\Gamma^{\l,pe}_{{\vec k}hee} = \frac{b_W}{\pi}G_{\rm F}^2n_n(|{\vec k}|+\Delta m)^2\delta_{h,-1},
\end{equation}
which is twice as large as the rate which damps the off-diagonals $\smash{{\hat f}^\l_{{\vec k}h\alpha e}}$ and $\smash{{\hat f}^\l_{{\vec k}he\beta}}$ and that the distributions $\smash{{\hat f}^\l_{{\vec k}h\alpha\beta}}$ with both $\alpha,\beta \neq e$ are not affected by this process at all. This is all in accordance with the expectations based on the measurement theory and also with the active-sterile mixing case studied in the previous section.

%
\subsection{Neutrino-nucleon scattering}
%

The evaluation of the $n\nu\leftrightarrow n\nu$ collision integral proceeds in a very similar fashion, except that now also the back-scattering term contains the coherent distribution functions. 
We again go to the rest-frame of the initial neutron and neglect the nucleon momenta and neutrino masses in kinematic relations. Then, after a little algebra, the collision integral becomes 
\begin{align}
\bar C^{\l,n\nu}_{{\vec k}hij} \approx 
-\frac{1}{4(2\pi)^5}G_{\rm F}^2  \, \delta_{h, -1} 
\smash{\sum_{l_1,l_1'}}
&\int {\rm d}^3p_{n_i}{\rm d}^3p_{p_\nu}{\rm d}^3p_{n_f}\delta(|\vec{k}|-|\vec{p}_\nu|)\delta^3({\vec p}_{n_f}-{\vec k} + {\vec p}_{\nu}) 
\nonumber\\
& \times 
\Big(\bar U_{il_1} \bar U_{l_1'l} m_n^2 |{\vec k}|^2 b_{\rm Z} (1 - a_{\rm Z} \hat {\vec k} \cdot \hat {\vec p}_\nu)
\nonumber\\
&\hskip 0.7cm\times (f^{\l}_{{\bm p}_{n_i}} f^{\l}_{{\bm k}hlj}\delta_{l_1,l_1'} -  
f^{\l}_{{\bm p}_{n_f}} f^{\l}_{{\bm p}_\nu h l_1l_1'} \delta_{lj} \big) 
+ (h.c.)_{i\leftrightarrow j} \Big),
\label{eq:nnu_first_version}
\end{align}
where we also made the MB-approximation. In the MB-limit we can rewrite the final state neutron (equilibrium) density as $\smash{f^<_{{\vec p}_{n_f}} = f^<_{{\vec p}_{n_i}}f^<_{{\vec k},{\rm eq}}/f^<_{{\vec p}_\nu,{\rm eq}}}=f^<_{{\vec p}_{n_i}}$, where the last equality follows from the kinematic constraint that restricts the $\nu$-scattering (off a heavy neutron target) to the elastic limit. This allows us to integrate over both neutron momenta in~\cref{eq:nnu_first_version}  and eventually also over the final state neutrino momentum, apart from angular coordinates. After these operations we can eventually write the collision integral in matrix notation as follows:
\begin{align}
\bar C^{\l,n\nu}_{{\vec k}h} = -\frac{b_Z}{2\pi} G_{\rm F}^2 n_n |{\vec k}|^2 \delta_{h,-1}
 \Big( \frac{1}{2}\{\bar U, f^{\l}_{{\bm k}h}\} 
 - \bar U \big\langle f^\l_{|{\vec k}|\Omega_{\vec k} h} \big\rangle \bar U 
 - a_Z\bar U \big\langle {\hat {\vec k}}\cdot {\hat {\vec p}}_\nu f^\l_{|{\vec k}|\Omega_{\vec k} h} \big\rangle \bar U  \Big),
\label{eq:nnu_second_version}
\end{align}
where curly brackets indicate an anti-commutator. To get~\cref{eq:nnu_second_version} we also used the idempotence $\bar U^2 = \bar U$ and Hermiticity $\bar U^\dagger = \bar U$ properties and defined the angular averaged moment functions
\begin{equation}
\big\langle X f^\l_{|{\vec k}|\Omega_{\vec k} h} \big\rangle \equiv \frac{1}{4\pi}\int {\rm d}\Omega_{\vec k} \,X f^\l_{|{\vec k}|\Omega_{\vec k} h}.
\end{equation}
The index structure of the final state neutrino distribution $f^\l_{|{\vec k}|\Omega_{\vec k} h}$ indicates that the state has the same energy and helicity $h=-1$ as the initial state neutrino, but in general has a different directional dependence. The flavour basis expression for the collision integral can be obtained from~\cref{eq:nnu_second_version} just by replacing $\bar U \rightarrow \hat U$. In the purely active-active mixing case, where $\hat U_{\alpha\beta} = \delta_{\alpha\beta}$ and $\bar U_{ij} = \delta_{ij}$ one then finds:
\begin{equation}
\bar C^{\l,n\nu}_{{\vec k}h,\rm active} = -\frac{b_Z}{2\pi} G_{\rm F}^2 n_n |{\vec k}|^2 \delta_{h,-1}
 \Big( \hat f^{\l}_{{\bm k}h} 
      - \big\langle \hat f^\l_{|{\vec k}|\Omega_{\vec k} h} \big\rangle
  - a_Z \big\langle {\hat {\vec k}}\cdot {\hat {\vec p}}_\nu \hat f^\l_{|{\vec k}|\Omega_{\vec k} h} \big\rangle   \Big).
\label{eq:nnu_third_version}
\end{equation}

It is evident from both~\cref{eq:nnu_second_version,eq:nnu_third_version} that the effect of the $n\nu \leftrightarrow n\nu$ collisions is strongly sensitive on the directional dependence of the distribution functions. Neutrinos in supernovae are indeed expected to have a nontrivial angular distributions, which also depend on the radial distance. At the very center the neutrino flux should be isotropic, while  at large distances beyond the neutrino sphere it would be enhanced in outgoing directions. The above formulae take this nontrivial dependence correctly into account.

\paragraph{Isotropic limit}

The expression~\cref{eq:nnu_third_version} clearly vanishes in the isotropic limit, where $f^{\l}_{{\bm k}h} = \langle f^\l_{|{\vec k}|\Omega_{\vec k}h} \rangle$ and $\langle {\hat {\vec k}}\cdot {\hat {\vec p}}_\nu f^\l_{|{\vec k}|\Omega_{\vec k}h} \rangle = 0$. The $Z$-boson mediated elastic scattering processes thus lose all resolution power on the flavour in the isotropic active-active mixing case, as was also observed in~\cite{Richers:2019grc}. In the active-sterile mixing case they do retain some resolution power even in the isotropic limit however. Indeed, using \eg~$\smash{(\hat U f^\l)_{\alpha\beta} = \delta_{\alpha a}f^\l_{a\beta}}$, one finds that in the isotropic limit active-sterile collision integral becomes
\begin{equation}
\bar C^{\l,n\nu, \rm a-s}_{{\vec k}h,{\rm isotropic},\alpha\beta} 
= -\frac{b_Z}{4\pi} G_{\rm F}^2 n_n |{\vec k}|^2 \delta_{h,-1}
 \Big( \delta_{\alpha a} \hat f^{\l}_{{\bm k}h a\beta  } 
     + \delta_{a\beta  } \hat f^{\l}_{{\bm k}h \alpha a}
     - 2 \delta_{\alpha a} \delta_{a\beta} \hat f^\l_{{\vec k}hab} \Big).
\label{eq:nnu_fourth_version}
\end{equation}
This expression resembles~\cref{eq:pe_third_version}, except that the back-scattering terms are now full isotropic distributions instead of the equilibrium distributions. As a result, the active-sterile $\smash{n\nu}$ collisions do damp the active-sterile flavour coherences, but they do not affect the active or sterile number density functions in the diagonal of $\smash{\hat f^{\l}_{{\bm k}h}}$ in the isotropic limit. \\

The UR-limit equations~\cref{eq:AH_UR}, written either in the mass eigenbasis or in the flavour basis, augmented with the exhaustive set of collision integrals can be used to describe neutrino flavour evolution all the way from the inside of the supernova to the detector at Earth, and to study non-trivial coherence effects during the whole process. However, this is out of the scope of this paper, whose sole purpose is to demonstrate the use of our formalism.
%
%
%
\section{Conclusions}
\label{sec:Conclusions}
%

We have presented quantum kinetic equations (QKE's) for coherently mixing neutrinos together with Feynman rules for computing collision integrals involving the coherent states. The neutrino QKE's were derived in more detail in the companion paper~\cite{Kainulainen:2023ocv}, but we gave a more rigorous treatment of the collision integrals and the derivation of the Feynman rules here than we did in~\cite{Kainulainen:2023ocv}. In particular, we showed rigorously that the extra overall phase factors in the forward scattering and collision terms cancel to all orders in perturbation theory. This proof is essential for the validity of our local, generalized Feynman rules. We changed slightly our notation in comparison to~\cite{Kainulainen:2023ocv}, related to the orientation of fermion lines in loop diagrams. While all results remain unchanged, the new convention is better in line with the common practice in field theory (that fermion lines are read opposite to the fermion flow), which may help to avoid some confusion.

The relevance of collision integrals encompassing local coherence effects has been emphasized more and more recently~\cite{Volpe:2023met}, but until now no comprehensive treatment existed for them, with the most general mixing structure valid for arbitrary neutrino masses and kinematics. For example in~\cite{McKellar:1992ja,Sigl:1992fn,Bennett:2020zkv,Froustey:2020mcq,Froustey:2021azz,Xiong:2022vsy,Hansen:2022xza, Capozzi:2018clo,Akita:2021hqn} the collision terms are computed ignoring the particle-antiparticle coherences, and in~\cite{Her11,Fid11,Jukkala:2019slc} particle-antiparticle coherence was included but without explicit expressions for the forward scattering and/or collision terms. In~\cite{Vlasenko:2013fja,Blaschke:2016xxt} collision and forward scattering terms are presented in the relativistic limit with helicity coherence but without particle-antiparticle coherence, and the analysis is based on expansions around small parameters. Our work here and in~\cite{Kainulainen:2023ocv} provides a systematic derivation of neutrino QKE's with all flavour and particle-antiparticle coherence effects valid for both light and heavy neutrinos.

Due to the generality of our formalism we can make firm remarks about the relevance of particle-antiparticle coherence. Our results in section~\cref{sec:flavour_and_particle-antiparticle_coherence} imply that neutrino-antineutrino coherences completely average out from the QKE's in essentially all neutrino physics setups, including heavy neutrinos in collision experiments. In particular, no effects are expected from this sector that could modify neutrino flavour evolution in supernovae. Neutrino-neutrino scattering processes have a rich flavour structure even in the absence of particle-antiparticle mixing however, and these effects can be relevant in dense environments. Our results, when applied in the UR-limit, are in full agreement with the earlier literature on this subject~\cite{Sigl:1992fn,Stirner:2018ojk,Fiorillo:2024fnl}.

We demonstrated the use of our formalism and Feynman rules by several worked out examples in section~\cref{sec:examples,sec:supernova}. Section~\cref{sec:examples} partly overlaps with the discussion given in~\cite{Kainulainen:2023ocv}, but we provide more details of the calculation here. In section~\cref{sec:supernova} we computed self-energy corrections and collision integrals relevant for arbitrary neutrino flavour mixing setup in supernovae. In particular, we gave explicit, remarkably simple expressions for the collision integrals for the $n\nu\leftrightarrow pe$ and $n\nu\leftrightarrow n\nu$ processes. Our QKE's correctly describe the evolution of the neutrino density matrix from the supernova to Earth and our formalism allows computing the resulting detector signals including flavour coherence and flavour separation effects. Other interesting applications of our formalism include (resonant) leptogenesis~\cite{Juk21} and the (p)reheating calculations at the end of inflation~\cite{Kainulainen:2021eki,Kainulainen:2022lzp,Kainulainen:2024etd}, as well as electroweak baryogenesis~\cite{Kai21}, with some modifications that will be discussed elsewhere.

%
\section*{Acknowledgements}
%

The work of HP was supported by grants from the Jenny and Antti Wihuri Foundation. He also thanks the Galileo Galilei Institute for Theoretical Physics for the hospitality and the INFN for partial support during the completion of this work.

%
\bibliography{references.bib}

\providecommand{\href}[2]{#2}\begingroup\raggedright\begin{thebibliography}{10}

\bibitem{Barbieri:1989ti}
R.~Barbieri and A.~Dolgov, \emph{{Bounds on Sterile-neutrinos from Nucleosynthesis}}, \href{https://doi.org/10.1016/0370-2693(90)91203-N}{\emph{Phys. Lett. B} {\bfseries 237} (1990) 440}.

\bibitem{Kainulainen:1990ds}
K.~Kainulainen, \emph{{Light Singlet Neutrinos and the Primordial Nucleosynthesis}}, \href{https://doi.org/10.1016/0370-2693(90)90054-A}{\emph{Phys. Lett. B} {\bfseries 244} (1990) 191}.

\bibitem{Barbieri:1990vx}
R.~Barbieri and A.~Dolgov, \emph{{Neutrino oscillations in the early universe}}, \href{https://doi.org/10.1016/0550-3213(91)90396-F}{\emph{Nucl. Phys. B} {\bfseries 349} (1991) 743}.

\bibitem{Enqvist:1990ad}
K.~Enqvist, K.~Kainulainen and J.~Maalampi, \emph{{Refraction and Oscillations of Neutrinos in the Early Universe}}, \href{https://doi.org/10.1016/0550-3213(91)90397-G}{\emph{Nucl. Phys. B} {\bfseries 349} (1991) 754}.

\bibitem{Enqvist:1990ek}
K.~Enqvist, K.~Kainulainen and J.~Maalampi, \emph{{Resonant neutrino transitions and nucleosynthesis}}, \href{https://doi.org/10.1016/0370-2693(90)91030-F}{\emph{Phys. Lett. B} {\bfseries 249} (1990) 531}.

\bibitem{Enqvist:1991qj}
K.~Enqvist, K.~Kainulainen and M.J.~Thomson, \emph{{Stringent cosmological bounds on inert neutrino mixing}}, \href{https://doi.org/10.1016/0550-3213(92)90442-E}{\emph{Nucl. Phys. B} {\bfseries 373} (1992) 498}.

\bibitem{Sigl:1992fn}
G.~Sigl and G.~Raffelt, \emph{{General kinetic description of relativistic mixed neutrinos}}, \href{https://doi.org/10.1016/0550-3213(93)90175-O}{\emph{Nucl. Phys. B} {\bfseries 406} (1993) 423}.

\bibitem{McKellar:1992ja}
B.H.J.~McKellar and M.J.~Thomson, \emph{{Oscillating doublet neutrinos in the early universe}}, \href{https://doi.org/10.1103/PhysRevD.49.2710}{\emph{Phys. Rev. D} {\bfseries 49} (1994) 2710}.

\bibitem{Volpe:2023met}
M.C.~Volpe, \emph{{Neutrinos from dense: flavor mechanisms, theoretical approaches, observations, new directions}},  \href{https://arxiv.org/abs/2301.11814}{{\ttfamily 2301.11814}}.

\bibitem{Volpe:2013uxl}
C.~Volpe, D.~V\"a\"an\"anen and C.~Espinoza, \emph{{Extended evolution equations for neutrino propagation in astrophysical and cosmological environments}}, \href{https://doi.org/10.1103/PhysRevD.87.113010}{\emph{Phys. Rev. D} {\bfseries 87} (2013) 113010} [\href{https://arxiv.org/abs/1302.2374}{{\ttfamily 1302.2374}}].

\bibitem{Vaananen:2013qja}
D.~V\"a\"an\"anen and C.~Volpe, \emph{{Linearizing neutrino evolution equations including neutrino-antineutrino pairing correlations}}, \href{https://doi.org/10.1103/PhysRevD.88.065003}{\emph{Phys. Rev. D} {\bfseries 88} (2013) 065003} [\href{https://arxiv.org/abs/1306.6372}{{\ttfamily 1306.6372}}].

\bibitem{Serreau:2014cfa}
J.~Serreau and C.~Volpe, \emph{{Neutrino-antineutrino correlations in dense anisotropic media}}, \href{https://doi.org/10.1103/PhysRevD.90.125040}{\emph{Phys. Rev. D} {\bfseries 90} (2014) 125040} [\href{https://arxiv.org/abs/1409.3591}{{\ttfamily 1409.3591}}].

\bibitem{Kartavtsev:2015eva}
A.~Kartavtsev, G.~Raffelt and H.~Vogel, \emph{{Neutrino propagation in media: Flavor-, helicity-, and pair correlations}}, \href{https://doi.org/10.1103/PhysRevD.91.125020}{\emph{Phys. Rev. D} {\bfseries 91} (2015) 125020} [\href{https://arxiv.org/abs/1504.03230}{{\ttfamily 1504.03230}}].

\bibitem{Her081}
M.~Herranen, K.~Kainulainen and P.M.~Rahkila, \emph{{Quantum kinetic theory for fermions in temporally varying backgrounds}}, \href{https://doi.org/10.1088/1126-6708/2008/09/032}{\emph{JHEP} {\bfseries 09} (2008) 032} [\href{https://arxiv.org/abs/0807.1435}{{\ttfamily 0807.1435}}].

\bibitem{Her083}
M.~Herranen, K.~Kainulainen and P.M.~Rahkila, \emph{{Kinetic theory for scalar fields with nonlocal quantum coherence}}, \href{https://doi.org/10.1088/1126-6708/2009/05/119}{\emph{JHEP} {\bfseries 05} (2009) 119} [\href{https://arxiv.org/abs/0812.4029}{{\ttfamily 0812.4029}}].

\bibitem{Her09}
M.~Herranen, K.~Kainulainen and P.M.~Rahkila, \emph{{Coherent quasiparticle approximation (cQPA) and nonlocal coherence}}, \href{https://doi.org/10.1088/1742-6596/220/1/012007}{\emph{J. Phys. Conf. Ser.} {\bfseries 220} (2010) 012007} [\href{https://arxiv.org/abs/0912.2490}{{\ttfamily 0912.2490}}].

\bibitem{Her10}
M.~Herranen, K.~Kainulainen and P.M.~Rahkila, \emph{{Coherent quantum Boltzmann equations from cQPA}}, \href{https://doi.org/10.1007/JHEP12(2010)072}{\emph{JHEP} {\bfseries 12} (2010) 072} [\href{https://arxiv.org/abs/1006.1929}{{\ttfamily 1006.1929}}].

\bibitem{Her11}
M.~Herranen, K.~Kainulainen and P.M.~Rahkila, \emph{{Flavour-coherent propagators and Feynman rules: Covariant cQPA formulation}}, \href{https://doi.org/10.1007/JHEP02(2012)080}{\emph{JHEP} {\bfseries 02} (2012) 080} [\href{https://arxiv.org/abs/1108.2371}{{\ttfamily 1108.2371}}].

\bibitem{Fid11}
C.~Fidler, M.~Herranen, K.~Kainulainen and P.M.~Rahkila, \emph{{Flavoured quantum Boltzmann equations from cQPA}}, \href{https://doi.org/10.1007/JHEP02(2012)065}{\emph{JHEP} {\bfseries 02} (2012) 065} [\href{https://arxiv.org/abs/1108.2309}{{\ttfamily 1108.2309}}].

\bibitem{Jukkala:2019slc}
H.~Jukkala, K.~Kainulainen and O.~Koskivaara, \emph{{Quantum transport and the phase space structure of the Wightman functions}}, \href{https://doi.org/10.1007/JHEP01(2020)012}{\emph{JHEP} {\bfseries 01} (2020) 012} [\href{https://arxiv.org/abs/1910.10979}{{\ttfamily 1910.10979}}].

\bibitem{Juk21}
H.~Jukkala, K.~Kainulainen and P.M.~Rahkila, \emph{{Flavour mixing transport theory and resonant leptogenesis}}, \href{https://doi.org/10.1007/JHEP09(2021)119}{\emph{JHEP} {\bfseries 09} (2021) 119} [\href{https://arxiv.org/abs/2104.03998}{{\ttfamily 2104.03998}}].

\bibitem{Kainulainen:2023ocv}
K.~Kainulainen and H.~Parkkinen, \emph{{Quantum transport theory for neutrinos with flavor and particle-antiparticle mixing}}, \href{https://doi.org/10.1007/JHEP02(2024)217}{\emph{JHEP} {\bfseries 02} (2024) 217} [\href{https://arxiv.org/abs/2309.00881}{{\ttfamily 2309.00881}}].

\bibitem{Sch61}
J.~Schwinger, \emph{Brownian motion of a quantum oscillator}, \href{https://doi.org/10.1063/1.1703727}{\emph{Journal of Mathematical Physics} {\bfseries 2} (1961) 407}.

\bibitem{Kel64}
L.~Keldysh, \emph{{Diagram technique for nonequilibrium processes}}, {\emph{Zh. Eksp. Teor. Fiz.} {\bfseries 47} (1964) 1515}.

\bibitem{Cal88}
E.~Calzetta and B.L.~Hu, \emph{Nonequilibrium quantum fields: Closed-time-path effective action, wigner function, and boltzmann equation}, \href{https://doi.org/10.1103/PhysRevD.37.2878}{\emph{Phys. Rev. D} {\bfseries 37} (1988) 2878}.

\bibitem{Cho85}
K.~chao Chou, Z.~bin Su, B.~lin Hao and L.~Yu, \emph{Equilibrium and nonequilibrium formalisms made unified}, \href{https://doi.org/https://doi.org/10.1016/0370-1573(85)90136-X}{\emph{Physics Reports} {\bfseries 118} (1985) 1 }.

\bibitem{Lut60}
J.M.~Luttinger and J.C.~Ward, \emph{Ground-state energy of a many-fermion system. ii}, \href{https://doi.org/10.1103/PhysRev.118.1417}{\emph{Phys. Rev.} {\bfseries 118} (1960) 1417}.

\bibitem{Cor74}
J.M.~Cornwall, R.~Jackiw and E.~Tomboulis, \emph{{Effective Action for Composite Operators}}, \href{https://doi.org/10.1103/PhysRevD.10.2428}{\emph{Phys. Rev. D} {\bfseries 10} (1974) 2428}.

\bibitem{Her082}
M.~Herranen, K.~Kainulainen and P.M.~Rahkila, \emph{{Towards a kinetic theory for fermions with quantum coherence}}, \href{https://doi.org/10.1016/j.nuclphysb.2008.09.032}{\emph{Nucl. Phys. B} {\bfseries 810} (2009) 389} [\href{https://arxiv.org/abs/0807.1415}{{\ttfamily 0807.1415}}].

\bibitem{Juk22}
H.~Jukkala, \emph{{Quantum coherence in relativistic transport theory : applications to baryogenesis}}, Ph.D. thesis, Jyvaskyla U., 2022.

\bibitem{Herranen:2010mh}
M.~Herranen, K.~Kainulainen and P.M.~Rahkila, \emph{{Coherent quantum Boltzmann equations from cQPA}}, \href{https://doi.org/10.1007/JHEP12(2010)072}{\emph{JHEP} {\bfseries 12} (2010) 072} [\href{https://arxiv.org/abs/1006.1929}{{\ttfamily 1006.1929}}].

\bibitem{Sawyer:2022ugt}
R.F.~Sawyer, \emph{{Neutrino-anti-neutrino instability in dense neutrino systems, with applications to the early universe and to supernovae}},  \href{https://arxiv.org/abs/2206.09290}{{\ttfamily 2206.09290}}.

\bibitem{Sawyer:2023dov}
R.F.~Sawyer, \emph{{Fast flavor evolution in dense neutrino systems, as described in quantum field theory}}, \href{https://doi.org/10.1103/PhysRevD.108.093001}{\emph{Phys. Rev. D} {\bfseries 108} (2023) 093001} [\href{https://arxiv.org/abs/2304.01987}{{\ttfamily 2304.01987}}].

\bibitem{Fiorillo:2024wej}
D.F.G.~Fiorillo, G.G.~Raffelt and G.~Sigl, \emph{{Collective neutrino-antineutrino oscillations in dense neutrino environments?}}, \href{https://doi.org/10.1103/PhysRevD.109.043031}{\emph{Phys. Rev. D} {\bfseries 109} (2024) 043031} [\href{https://arxiv.org/abs/2401.02478}{{\ttfamily 2401.02478}}].

\bibitem{Stirner:2018ojk}
T.~Stirner, G.~Sigl and G.~Raffelt, \emph{{Liouville term for neutrinos: Flavor structure and wave interpretation}}, \href{https://doi.org/10.1088/1475-7516/2018/05/016}{\emph{JCAP} {\bfseries 05} (2018) 016} [\href{https://arxiv.org/abs/1803.04693}{{\ttfamily 1803.04693}}].

\bibitem{Fiorillo:2024fnl}
D.F.G.~Fiorillo, G.G.~Raffelt and G.~Sigl, \emph{{Inhomogeneous Kinetic Equation for Mixed Neutrinos: Tracing the Missing Energy}}, \href{https://doi.org/10.1103/PhysRevLett.133.021002}{\emph{Phys. Rev. Lett.} {\bfseries 133} (2024) 021002} [\href{https://arxiv.org/abs/2401.05278}{{\ttfamily 2401.05278}}].

\bibitem{Xiong:2022vsy}
Z.~Xiong, M.-R.~Wu, G.~Mart\'\i{}nez-Pinedo, T.~Fischer, M.~George, C.-Y.~Lin et~al., \emph{{Evolution of collisional neutrino flavor instabilities in spherically symmetric supernova models}}, \href{https://doi.org/10.1103/PhysRevD.107.083016}{\emph{Phys. Rev. D} {\bfseries 107} (2023) 083016} [\href{https://arxiv.org/abs/2210.08254}{{\ttfamily 2210.08254}}].

\bibitem{Lin:2022dek}
Y.-C.~Lin and H.~Duan, \emph{{Collision-induced flavor instability in dense neutrino gases with energy-dependent scattering}}, \href{https://doi.org/10.1103/PhysRevD.107.083034}{\emph{Phys. Rev. D} {\bfseries 107} (2023) 083034} [\href{https://arxiv.org/abs/2210.09218}{{\ttfamily 2210.09218}}].

\bibitem{PhysRevD.106.103029}
L.~Johns and Z.~Xiong, \emph{Collisional instabilities of neutrinos and their interplay with fast flavor conversion in compact objects}, \href{https://doi.org/10.1103/PhysRevD.106.103029}{\emph{Phys. Rev. D} {\bfseries 106} (2022) 103029}.

\bibitem{PhysRevD.106.103031}
I.~Padilla-Gay, I.~Tamborra and G.G.~Raffelt, \emph{Neutrino fast flavor pendulum. ii. collisional damping}, \href{https://doi.org/10.1103/PhysRevD.106.103031}{\emph{Phys. Rev. D} {\bfseries 106} (2022) 103031}.

\bibitem{Ehring:2023lcd}
J.~Ehring, S.~Abbar, H.-T.~Janka, G.~Raffelt and I.~Tamborra, \emph{{Fast neutrino flavor conversion in core-collapse supernovae: A parametric study in 1D models}}, \href{https://doi.org/10.1103/PhysRevD.107.103034}{\emph{Phys. Rev. D} {\bfseries 107} (2023) 103034} [\href{https://arxiv.org/abs/2301.11938}{{\ttfamily 2301.11938}}].

\bibitem{Tamborra:2017ubu}
I.~Tamborra, L.~Huedepohl, G.~Raffelt and H.-T.~Janka, \emph{{Flavor-dependent neutrino angular distribution in core-collapse supernovae}}, \href{https://doi.org/10.3847/1538-4357/aa6a18}{\emph{Astrophys. J.} {\bfseries 839} (2017) 132} [\href{https://arxiv.org/abs/1702.00060}{{\ttfamily 1702.00060}}].

\bibitem{Tamborra:2020cul}
I.~Tamborra and S.~Shalgar, \emph{{New Developments in Flavor Evolution of a Dense Neutrino Gas}}, \href{https://doi.org/10.1146/annurev-nucl-102920-050505}{\emph{Ann. Rev. Nucl. Part. Sci.} {\bfseries 71} (2021) 165} [\href{https://arxiv.org/abs/2011.01948}{{\ttfamily 2011.01948}}].

\bibitem{Johns:2019izj}
L.~Johns, H.~Nagakura, G.M.~Fuller and A.~Burrows, \emph{{Neutrino oscillations in supernovae: angular moments and fast instabilities}}, \href{https://doi.org/10.1103/PhysRevD.101.043009}{\emph{Phys. Rev. D} {\bfseries 101} (2020) 043009} [\href{https://arxiv.org/abs/1910.05682}{{\ttfamily 1910.05682}}].

\bibitem{Capozzi:2018clo}
F.~Capozzi, B.~Dasgupta, A.~Mirizzi, M.~Sen and G.~Sigl, \emph{{Collisional triggering of fast flavor conversions of supernova neutrinos}}, \href{https://doi.org/10.1103/PhysRevLett.122.091101}{\emph{Phys. Rev. Lett.} {\bfseries 122} (2019) 091101} [\href{https://arxiv.org/abs/1808.06618}{{\ttfamily 1808.06618}}].

\bibitem{Richers:2019grc}
S.A.~Richers, G.C.~McLaughlin, J.P.~Kneller and A.~Vlasenko, \emph{{Neutrino Quantum Kinetics in Compact Objects}}, \href{https://doi.org/10.1103/PhysRevD.99.123014}{\emph{Phys. Rev. D} {\bfseries 99} (2019) 123014} [\href{https://arxiv.org/abs/1903.00022}{{\ttfamily 1903.00022}}].

\bibitem{Vlasenko:2013fja}
A.~Vlasenko, G.M.~Fuller and V.~Cirigliano, \emph{{Neutrino Quantum Kinetics}}, \href{https://doi.org/10.1103/PhysRevD.89.105004}{\emph{Phys. Rev. D} {\bfseries 89} (2014) 105004} [\href{https://arxiv.org/abs/1309.2628}{{\ttfamily 1309.2628}}].

\bibitem{Bennett:2020zkv}
J.J.~Bennett, G.~Buldgen, P.F.~De~Salas, M.~Drewes, S.~Gariazzo, S.~Pastor et~al., \emph{{Towards a precision calculation of $N_{\rm eff}$ in the Standard Model II: Neutrino decoupling in the presence of flavour oscillations and finite-temperature QED}}, \href{https://doi.org/10.1088/1475-7516/2021/04/073}{\emph{JCAP} {\bfseries 04} (2021) 073} [\href{https://arxiv.org/abs/2012.02726}{{\ttfamily 2012.02726}}].

\bibitem{Froustey:2020mcq}
J.~Froustey, C.~Pitrou and M.C.~Volpe, \emph{{Neutrino decoupling including flavour oscillations and primordial nucleosynthesis}}, \href{https://doi.org/10.1088/1475-7516/2020/12/015}{\emph{JCAP} {\bfseries 12} (2020) 015} [\href{https://arxiv.org/abs/2008.01074}{{\ttfamily 2008.01074}}].

\bibitem{Froustey:2021azz}
J.~Froustey and C.~Pitrou, \emph{{Primordial neutrino asymmetry evolution with full mean-field effects and collisions}}, \href{https://doi.org/10.1088/1475-7516/2022/03/065}{\emph{JCAP} {\bfseries 03} (2022) 065} [\href{https://arxiv.org/abs/2110.11889}{{\ttfamily 2110.11889}}].

\bibitem{Hansen:2022xza}
R.S.L.~Hansen, S.~Shalgar and I.~Tamborra, \emph{{Enhancement or damping of fast neutrino flavor conversions due to collisions}}, \href{https://doi.org/10.1103/PhysRevD.105.123003}{\emph{Phys. Rev. D} {\bfseries 105} (2022) 123003} [\href{https://arxiv.org/abs/2204.11873}{{\ttfamily 2204.11873}}].

\bibitem{Akita:2021hqn}
K.~Akita, G.~Lambiase and M.~Yamaguchi, \emph{{Unstable cosmic neutrino capture}}, \href{https://doi.org/10.1007/JHEP02(2022)132}{\emph{JHEP} {\bfseries 02} (2022) 132} [\href{https://arxiv.org/abs/2109.02900}{{\ttfamily 2109.02900}}].

\bibitem{Blaschke:2016xxt}
D.N.~Blaschke and V.~Cirigliano, \emph{{Neutrino Quantum Kinetic Equations: The Collision Term}}, \href{https://doi.org/10.1103/PhysRevD.94.033009}{\emph{Phys. Rev. D} {\bfseries 94} (2016) 033009} [\href{https://arxiv.org/abs/1605.09383}{{\ttfamily 1605.09383}}].

\bibitem{Kainulainen:2021eki}
K.~Kainulainen and O.~Koskivaara, \emph{{Non-equilibrium dynamics of a scalar field with quantum backreaction}}, \href{https://doi.org/10.1007/JHEP12(2021)190}{\emph{JHEP} {\bfseries 12} (2021) 190} [\href{https://arxiv.org/abs/2105.09598}{{\ttfamily 2105.09598}}].

\bibitem{Kainulainen:2022lzp}
K.~Kainulainen, O.~Koskivaara and S.~Nurmi, \emph{{Tachyonic production of dark relics: a non-perturbative quantum study}},  \href{https://arxiv.org/abs/2209.10945}{{\ttfamily 2209.10945}}.

\bibitem{Kainulainen:2024etd}
K.~Kainulainen, S.~Nurmi and O.~V\"ais\"anen, \emph{{Tachyonic production of dark relics: classical lattice vs. quantum 2PI in Hartree truncation}},  \href{https://arxiv.org/abs/2406.17468}{{\ttfamily 2406.17468}}.

\bibitem{Kai21}
K.~Kainulainen, \emph{{CP-violating transport theory for electroweak baryogenesis with thermal corrections}}, \href{https://doi.org/10.1088/1475-7516/2021/11/042}{\emph{JCAP} {\bfseries 11} (2021) 042} [\href{https://arxiv.org/abs/2108.08336}{{\ttfamily 2108.08336}}].

\end{thebibliography}\endgroup

\end{document}